\documentclass[aps,prl,reprint,superscriptaddress,longbibliography,footinbib,floatfix]{revtex4-2}

\usepackage[T1]{fontenc}
\usepackage{amsmath,amssymb,amsthm}
\usepackage{txfonts}
\usepackage{bm}
\usepackage{graphicx}
\usepackage{xcolor}
\usepackage{pgfplots}
\pgfplotsset{compat=1.16}
\usetikzlibrary{decorations.pathmorphing}
\usepackage[expansion=false]{microtype}
\usepackage[colorlinks=true,allcolors=blue]{hyperref}
\usepackage{orcidlink}

\newtheorem{theorem}{Theorem}
\newtheorem{lemma}[theorem]{Lemma}
\newtheorem{proposition}[theorem]{Proposition}

\newcommand{\Tr}{\operatorname{Tr}}
\newcommand{\id}{\mathbb I}
\newcommand{\ket}[1]{|#1\rangle}
\newcommand{\bra}[1]{\langle#1|}
\newcommand{\ketbra}[2]{|#1\rangle\langle#2|}
\newcommand{\EF}{E_F}
\newcommand{\GF}{G_F}
\newcommand{\EW}{\mathcal E}
\newcommand{\Pee}{\Psi_{\rm ee}}
\newcommand{\Poo}{\Psi_{\rm oo}}

\newcommand{\SN}{\mathrm{SN}}
\newcommand{\papertitle}{Non-Optimality of Fermionic Gaussian Entanglement of Formation}

\begin{document}

\title{\papertitle}

\author{Gerardo Adesso\orcidlink{0000-0001-7136-3755}}
\email{gerardo.adesso@nottingham.ac.uk}
\affiliation{Algorithmiq S.r.L., Via della Chiusa 15, 20123 Milano, Italy}
\affiliation{School of Mathematical Sciences and Centre for Mathematics and
Theoretical Physics of Quantum Non-Equilibrium Systems,
University of Nottingham, University Park, Nottingham NG7 2RD, United Kingdom}

\date{\today}

\begin{abstract}
Gaussian states are a mainstay of bosonic and fermionic quantum physics alike. For bosons, the cheapest way to prepare a Gaussian state from pure states, as quantified by the entanglement of formation, has recently been proven to be Gaussian. Here we show that this fails for fermions. Mixing non-Gaussian pure states can produce a fermionic Gaussian state with strictly less entanglement than any mixture of Gaussian ones. Gaussian preparations are suboptimal for the R\'enyi-$\alpha$ entanglement of formation at all orders $\alpha\geq1$, as we demonstrate on two copies of a thermal Cooper pair state. The gap also appears in the Schmidt number: Gaussian decompositions can require pure states of higher Schmidt rank than general ones. Nevertheless, the pairing amplitude alone provides a lower bound on the entanglement of formation of any state with two modes per party.
\end{abstract}

\maketitle
\nocite{apsrev42Control}

Entanglement is a resource: two distant parties acting by local operations and classical communication (LOCC) must consume it whenever they assemble a quantum correlated state \cite{HorodeckiReview,PlenioVirmani}. The basic measure of this consumption is the entanglement of formation $\EF$, the least average entanglement of the pure states used in any preparation of the state \cite{BennettFormation}. Its regularization gives the asymptotic entanglement cost \cite{HaydenHorodeckiTerhal}, while the Schmidt number sets the zero-error one-shot cost of preparing a single copy \cite{TerhalHorodecki,BuscemiDatta}. Each of these quantities requires an optimization over all decompositions of the state. This is intractable in general, and closed forms are known only in special cases, like two qubits \cite{Wootters} and  highly symmetric families~\cite{VollbrechtWerner}.

Bosonic Gaussian states are a notable exception \cite{Weedbrook,SerafiniBook,ARLReview}. They describe coherent and squeezed light, thermal oscillators and much of continuous-variable quantum optics, and every mixed Gaussian state can be prepared by randomly displacing a single pure Gaussian state, which provides a natural Gaussian decomposition \cite{Wolf}. A long line of work has shown that no other preparation is cheaper. Gaussian decompositions were first proven optimal for symmetric two-mode states \cite{Giedke}, then for all two-mode states \cite{AdessoEoF} and for any number of modes \cite{GA}, and the same holds for the R\'enyi entanglement of formation at all orders above one \cite{GARenyi}. Computing the entanglement of formation of a bosonic Gaussian state thus reduces to a tractable optimization over covariance matrices.

Fermionic Gaussian states are just as central \cite{SuraceTagliacozzo,HacklBianchi}. In many-body physics they are the ground and thermal states of quadratic Hamiltonians, which include free fermions as well as the mean-field and BCS descriptions of superconductors, atomic nuclei and molecules \cite{BCS,KrausCirac,KrausPairing,DingSchilling,DingMolecules}. In quantum computing they are the states produced by free fermion (matchgate) circuits \cite{Valiant,TerhalDiVincenzo}. These circuits can be simulated efficiently on a classical computer \cite{JozsaMiyake,BravyiKoenig}, and become universal only when supplied with non-Gaussian ``magic'' states \cite{Hebenstreit}. It is therefore natural to ask whether for fermionic Gaussian states, too, Gaussian preparations are the cheapest.

The fermionic question is genuinely different. The parity superselection rule forbids superpositions of even and odd fermion number \cite{Zanardi,WisemanVaccaro,SchuchVerstraeteCirac,BanulsCiracWolf,FriisLeeBruschi,BenattiReview,SzalayFermionic}, which restricts the pure states allowed in a decomposition. A mixed fermionic Gaussian state can be decomposed into pure Gaussian states, for instance the occupation-number states of its normal modes, but these generally carry different amounts of entanglement. Salek \emph{et al.}\ recently defined the fermionic Gaussian entanglement of formation $\GF$, proved Gaussian optimality $\EF=\GF$ for states with at most three mixed normal modes and for several structured families, and left open whether non-Gaussian decompositions can ever do better, noting that they first become available with four mixed normal modes \cite{Salek}.

In this Letter we show that non-Gaussian decompositions of fermionic Gaussian states \emph{can} do better, requiring less entanglement than Gaussian ones, already in the smallest case left open: two modes on each side. The reason is structural. Superselection splits two modes per party into two virtual qubit pairs, and Gaussianity imposes an algebraic condition that ties the entanglement of one pair to that of the other (Lemma~\ref{lem:quadric}). A non-Gaussian decomposition can concentrate the entanglement in one pair and so use less of it overall. For two copies of a thermal Cooper pair state, we prove that Gaussian decompositions are strictly suboptimal for the R\'enyi entanglement of formation at every order $\alpha\geq1$ (Theorems~\ref{thm:analytic} and~\ref{thm:renyi}). The \emph{gap} also appears in the Schmidt number: Gaussian decompositions need pure states of Schmidt rank four, where rank three suffices in general (Theorem~\ref{thm:schmidt}). On the positive side, the pairing amplitude alone gives a lower bound on the entanglement of formation of every state with two modes per party (Theorem~\ref{thm:pairing}), a remnant of the bosonic covariance bounds.

\emph{Setting.}---A fermionic mode, such as an electronic orbital or a lattice site, can be empty or occupied by one fermion. For modes $c_1,\ldots,c_N$, the annihilation and creation operators $c_j$ and $c_j^\dagger$ obey the canonical anticommutation relations $\{c_j,c_k^\dagger\}=\delta_{jk}$ and $\{c_j,c_k\}=0$, the occupation numbers $n_j=c_j^\dagger c_j\in\{0,1\}$ label the Fock basis $\ket{n_1\cdots n_N}$, and the parity operator $P=(-1)^{\sum_jn_j}$ records whether the total number of fermions is even ($P=+1$) or odd ($P=-1$) \cite{SzalayFermionic,SuraceTagliacozzo}. Alice holds two modes $a_1,a_2$ and Bob holds two modes $b_1,b_2$, so that $(c_1,\ldots,c_4)=(a_1,a_2,b_1,b_2)$; for a single pair we write $a$ and $b$. It is convenient to work with the Hermitian Majorana operators $\gamma_{2j-1}=c_j+c_j^\dagger$ and $\gamma_{2j}=i(c_j^\dagger-c_j)$, which obey $\{\gamma_k,\gamma_l\}=2\delta_{kl}$ and give $P=\prod_j(-i\gamma_{2j-1}\gamma_{2j})$; products of Majorana operators form a natural operator basis, used for instance in the classical simulation of fermionic circuits \cite{MajoranaPropagation}. The second moments of a state $\rho$ form its real antisymmetric covariance matrix $M_{kl}=\frac i2\Tr(\rho[\gamma_k,\gamma_l])$, which collects the occupations $\langle n_j\rangle$, the hopping amplitudes $\langle c_j^\dagger c_k\rangle$ and the pairing amplitudes $\langle c_j^\dagger c_k^\dagger\rangle$. A state is Gaussian if it is the ground or thermal state of a Hamiltonian quadratic in the Majoranas, $\rho\propto\exp(-\frac i4\sum_{kl}h_{kl}\gamma_k\gamma_l)$ with $h$ real antisymmetric, or a limit of such states. It is then fixed by $M$ alone through Wick's theorem, and a pure state is Gaussian if and only if $MM^\top=\id$ \cite{BravyiLagrangian,SuraceTagliacozzo,HacklBianchi}. Unitaries generated by quadratic Hamiltonians are called Gaussian; they are the free fermion (matchgate) circuits \cite{TerhalDiVincenzo,JozsaMiyake} and map Gaussian states to Gaussian states.

Physical fermionic states commute with $P$: the parity superselection rule forbids coherent superpositions of even and odd fermion number, so local operations can neither create such superpositions nor add a single fermion coherently \cite{WisemanVaccaro,BanulsCiracWolf,FriisLeeBruschi,SzalayFermionic}. To compute reduced states, we map the modes to qubits by Jordan--Wigner transformation in the order $a_1,a_2,b_1,b_2$; with Alice's modes first, tracing out Bob's modes is then the ordinary partial trace over his qubits \cite{BanulsCiracWolf,FriisLeeBruschi}.

The entanglement of a pure state $\psi$ is measured by the entropy of Alice's reduced state $\psi_A$, whose eigenvalues are the Schmidt coefficients of $\psi$. We use the R\'enyi entropies
\begin{equation}
 S_\alpha(\psi_A)=\frac1{1-\alpha}\log\Tr\psi_A^\alpha,
\end{equation}
with logarithms to base $2$, so that entanglement is counted in ebits. They include the von Neumann entropy $S_1\equiv S$, as $\alpha\to1$, and the min-entropy $S_\infty(\psi_A)=-\log\|\psi_A\|_\infty$, as $\alpha\to\infty$, where $\|\psi_A\|_\infty$ is the largest Schmidt coefficient. For a mixed state, following Refs.~\cite{BanulsCiracWolf,Salek}, the R\'enyi-$\alpha$ entanglement of formation is the least average entanglement over decompositions,
\begin{equation}
 E_{F,\alpha}(\rho)=\inf_{\{p_i,\psi_i\}}\sum_ip_i\,S_\alpha(\psi_{i,A}),
 \label{eq:EF}
\end{equation}
where the infimum runs over decompositions $\rho=\sum_ip_i\psi_i$ into pure states of definite parity, as superselection requires, with probabilities $p_i$; at $\alpha=1$ this is the entanglement of formation $E_{F,1}\equiv\EF$ \cite{BennettFormation}. The Gaussian entanglement of formation $G_{F,\alpha}$ is defined analogously, but with the infimum restricted to pure Gaussian states $\psi_i$; it is finite on every Gaussian state \cite{Salek}, and $E_{F,\alpha}\le G_{F,\alpha}$, with $G_{F,1}\equiv\GF$. Since each $\psi_i$ has definite parity, decompositions of $\rho$ split into decompositions of its even and odd blocks $\rho_\pm$, so a nonzero gap $G_{F,\alpha}-E_{F,\alpha}$ on either block implies one for $\rho$; we will focus on the normalized even block $\hat\rho_+=\rho_+/\Tr\rho_+$.

\begin{figure}[tb]
\centering
\begin{tikzpicture}[x=1cm,y=1cm,font=\small]
 \node[font=\small\bfseries] at (0,2.35) {Alice};
 \node[font=\small\bfseries] at (4.2,2.35) {Bob};
 \draw[rounded corners,fill=blue!6,draw=blue!50] (-0.75,1.05) rectangle (0.75,1.95);
 \draw[rounded corners,fill=blue!6,draw=blue!50] (3.45,1.05) rectangle (4.95,1.95);
 \draw[rounded corners,fill=orange!8,draw=orange!60] (-0.75,-0.05) rectangle (0.75,0.85);
 \draw[rounded corners,fill=orange!8,draw=orange!60] (3.45,-0.05) rectangle (4.95,0.85);
 \node at (0,1.5) {$\ket{00},\ket{11}$};
 \node at (4.2,1.5) {$\ket{00},\ket{11}$};
 \node at (0,0.4) {$\ket{01},\ket{10}$};
 \node at (4.2,0.4) {$\ket{01},\ket{10}$};
 \draw[thick,blue!70,decorate,decoration={snake,amplitude=1.2pt,segment length=6pt}] (0.8,1.5) -- (3.4,1.5) node[midway,above=1pt,text=blue!70!black] {$\Pee$};
 \draw[thick,orange!80,decorate,decoration={snake,amplitude=1.2pt,segment length=6pt}] (0.8,0.4) -- (3.4,0.4) node[midway,below=1pt,text=orange!70!black] {$\Poo$};
 \node[anchor=east,font=\footnotesize] at (-0.8,1.5) {even};
 \node[anchor=east,font=\footnotesize] at (-0.8,0.4) {odd};
 \node[font=\footnotesize,align=center] at (2.1,-0.55) {Gaussian lock: $\det\Pee+\det\Poo=0$};
\end{tikzpicture}
\caption{\label{fig:pairs}Under parity superselection, an even pure state of two fermionic modes per party is a superposition of two virtual qubit pairs: one between the even subspaces of Alice and Bob, and one between their odd subspaces. Gaussian states obey a single constraint linking the two (Lemma~\ref{lem:quadric}); non-Gaussian states may entangle one pair and leave the other nearly untouched.}
\end{figure}
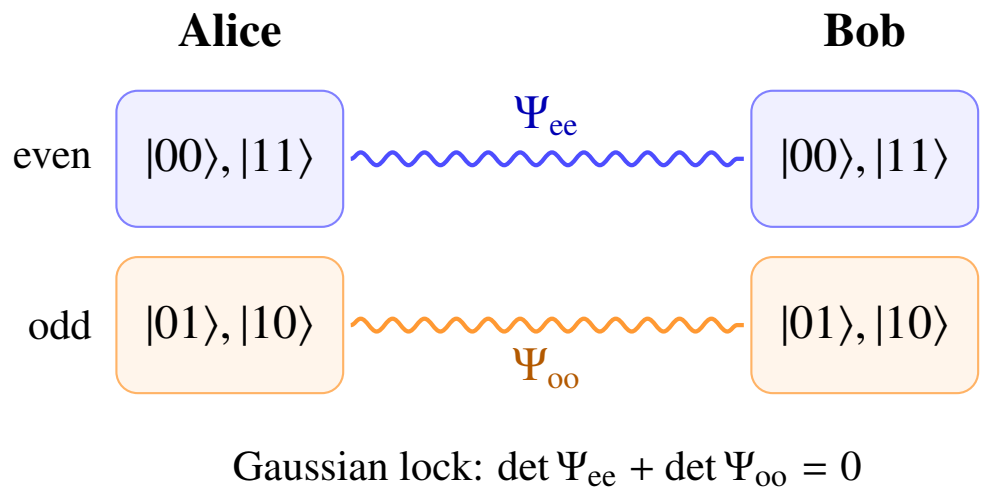

\emph{Two virtual qubit pairs.}---Write a parity-even pure state as $\ket\psi=\sum_{x,y}\Psi_{xy}\ket x_A\ket y_B$, with $x,y\in\{00,01,10,11\}$ the occupations of Alice's and Bob's modes. Parity forces $\Psi=\Pee\oplus\Poo$ to be block diagonal: $\Pee$ couples Alice's even states $\ket{00},\ket{11}$ to Bob's, and $\Poo$ couples their odd states $\ket{01},\ket{10}$ (Fig.~\ref{fig:pairs}). The state is therefore a superposition of two two-qubit states, which we call virtual qubit pairs. Let $\mathcal P=\|\Pee\|^2$ be the probability that Alice's parity is even, and let $C_{\rm ee}=2|\det\Pee|/\mathcal P$ and $C_{\rm oo}=2|\det\Poo|/(1-\mathcal P)$ be the concurrences of the normalized blocks. The entanglement of $\psi$ is then
\begin{equation}
 S(\psi_A)=h(\mathcal P)+\mathcal P\,\EW(C_{\rm ee})+(1-\mathcal P)\,\EW(C_{\rm oo}),
 \label{eq:Sblocks}
\end{equation}
where $h$ is the binary entropy and $\EW(C)=h\big(\frac12+\frac12\sqrt{1-C^2}\big)$ is the concurrence function of Wootters \cite{Wootters}.\footnote{A $2\times2$ positive matrix of trace $x$ and determinant $c^2$ has eigenvalues $\frac12(x\pm\sqrt{x^2-4c^2})$; applied to $\psi_A=\Pee\Pee^\dagger\oplus\Poo\Poo^\dagger$, this gives Eq.~\eqref{eq:Sblocks}.} The first term is carried by Alice's parity, and the other two are ordinary two-qubit entanglement.

\begin{lemma}[Gaussian lock]\label{lem:quadric}
Let $\psi$ be a parity-even pure state of two modes per party, with blocks $\Pee$ and $\Poo$. \\
(i) $\psi$ is Gaussian if and only if $\det\Pee+\det\Poo=0$. \\
(ii) Local Gaussian unitaries act as $\Pee\mapsto u_1\Pee v_1^{\top}$ and $\Poo\mapsto u_2\Poo v_2^{\top}$ with independent $u_k,v_k\in SU(2)$, and therefore leave both determinants unchanged.
\end{lemma} 

We prove Lemma~\ref{lem:quadric} in the End Matter. Part (i) is the Pfaffian identity obeyed by the vacuum, two-particle and four-particle amplitudes of a Gaussian state, rewritten in Jordan--Wigner signs, together with its converse: geometrically, even pure Gaussian states form exactly a quadric of pure spinors \cite{Chevalley,SarosiLevay}, the state counterpart of Valiant's matchgate identity \cite{Valiant,JozsaMiyake}. Part (ii) holds because Alice's local Gaussian group $\mathrm{Spin}(4)\simeq SU(2)\times SU(2)$ acts separately on her even and odd subspaces. Thus a Gaussian state must split its entanglement between the two virtual pairs in the fixed ratio $\mathcal P\,C_{\rm ee}=(1-\mathcal P)\,C_{\rm oo}$, whereas a general state can put all of it into one pair, and no local Gaussian unitary can change this. How strongly a state violates the lock can be quantified by the Williamson spectrum of the covariance matrix, i.e.\ the moduli of the eigenvalues of $M$, which all equal $1$ exactly for pure Gaussian states \cite{TurkeshiWilliamson}. Indeed, every even pure state of two modes per party has $MM^\top=(1-4|\det\Pee+\det\Poo|^2)\,\id$ (End Matter), in agreement with Ref.~\cite{GigenaRossignoli}. The lock has no bosonic counterpart, since without a parity rule the Hilbert space does not split into virtual pairs. It also explains the known positive results: with at most three mixed normal modes every parity-definite state in the support is Gaussian \cite{Salek}, and on a $1{:}3$ cut each block has rank one, so there is nothing to lock.

\emph{Cooper pairs.}---We demonstrate our findings on two Cooper pairs shared between Alice and Bob. Each pair $(a_j,b_j)$, with one mode on either side, is in the thermal state $\sigma\propto\exp[-d(a_j^\dagger b_j^\dagger+b_ja_j)]$ of a pairing Hamiltonian, with $d<0$, so that the four modes are in the state $\rho=\sigma\otimes\sigma$. The covariance matrix of $\rho$ is $t$ times that of the pure Gaussian state $\Phi^+\otimes\Phi^+$, where $t=\tanh(|d|/2)$ and $\ket{\Phi^+}=(\ket{00}+\ket{11})/\sqrt2$ is the Bell state of one Cooper pair. The local Gaussian unitaries that leave $\Phi^+\otimes\Phi^+$ invariant therefore leave $\rho$ invariant too; they form the group $SU(2)\times SU(2)$, one factor rotating each virtual pair. We call the states $\rho$, for $0<t<1$, the isotropic Cooper pair family. Twirling over the symmetry group maps any decomposition of $\rho$ to another with the same Schmidt coefficients, and Gaussian decompositions to Gaussian ones \cite{VollbrechtWerner,TerhalVollbrecht}.
 After twirling, a pure even state is described by three {counting statistics}: the parity probability $\mathcal P$, and the probabilities $\mathcal{A}=|\langle\Phi_{\rm ee}|\psi\rangle|^2$ and $\mathcal{B}=|\langle\Phi_{\rm oo}|\psi\rangle|^2$ of finding the two virtual pairs in their Bell states $\ket{\Phi_{\rm ee}}=(\ket{00}_A\ket{00}_B-\ket{11}_A\ket{11}_B)/\sqrt2$ and $\ket{\Phi_{\rm oo}}=(\ket{01}_A\ket{01}_B+\ket{10}_A\ket{10}_B)/\sqrt2$, with $\ket{\Phi^+}\otimes\ket{\Phi^+}=(\ket{\Phi_{\rm ee}}+\ket{\Phi_{\rm oo}})/\sqrt2$ once the modes are reordered as $a_1,a_2,b_1,b_2$. The twirl also retains the coherence $\langle\Phi_{\rm ee}|\psi\rangle\langle\psi|\Phi_{\rm oo}\rangle$ between the two Bell states, whose modulus $\sqrt{\mathcal A\mathcal B}$ is fixed. A decomposition of the normalized even block $\hat\rho_+$ is therefore a probability distribution over $(\mathcal P,\mathcal A,\mathcal B)$ with
\begin{equation}
 \mathbb E[\mathcal P]=\tfrac12,\quad \mathbb E[\mathcal A]=\mathbb E[\mathcal B]=a,\quad \mathbb E\sqrt{\mathcal A\mathcal B}\ge\kappa,
 \label{eq:counting}
\end{equation}
where $a\pm\kappa=(1\pm t)^4/[8(1+t^4)]$; the inequality holds because $\hat\rho_+$ fixes the average coherence to $\kappa$, which cannot exceed the average modulus. The entanglement of formation thus becomes a problem about distributions of three numbers, and Gaussianity constrains each element to have the Schmidt coefficients of two independent pairs, with $\mathcal A$ and $\mathcal B$ bounded accordingly (End Matter).

\emph{Non-optimality at every order.}---We are ready to show that, on the isotropic Cooper pair family, Gaussian decompositions are strictly suboptimal for the R\'enyi entanglement of formation at all orders $\alpha \geq 1$. The cleanest statement concerns the min-entropy, $\alpha=\infty$. Let $\Lambda=\sup\sum_ip_i\|\psi_{i,A}\|_\infty$ be the largest average leading Schmidt coefficient over decompositions of $\hat\rho_+$, and $\Lambda_G$ the same over Gaussian decompositions. By convexity of $-\log$, $G_{F,\infty}\ge-\log\Lambda_G$, while a flat decomposition, whose elements all share one reduced spectrum with largest eigenvalue $\lambda$, gives $E_{F,\infty}\le-\log\lambda$.

\begin{theorem}[Min-entropy]\label{thm:analytic}
For the isotropic Cooper pair state $\rho$ at every $t\in(0,1)$,
\begin{align}
 \Lambda_G&\le\frac{1-t^2+t^4+(1-t^2)\sqrt{1+t^4}}{2(1+t^4)},\\
 \Lambda&\ge\begin{cases}\dfrac{\big[1+t^2+\sqrt3\,(1-t^2)\big]^2}{8(1+t^4)}, & t^2\ge2-\sqrt3,\\[8pt] \dfrac{1-t^2+t^4}{1+t^4}, & t^2\le2-\sqrt3.\end{cases}
\end{align}
In both regimes $\Lambda>\Lambda_G$, and the lower bounds on $\Lambda$ are attained by flat decompositions, so $E_{F,\infty}(\rho)<G_{F,\infty}(\rho)$.
\end{theorem}

The better decomposition is a mirror pair of states in which one virtual pair is maximally entangled and the other is a product state or nearly so, together supplying the coherence that $\rho$ requires. The Gaussian bound takes a few lines: von Neumann's trace inequality bounds the Bell-state probabilities of a Gaussian element, Cauchy--Schwarz turns the coherence constraint into a constraint on the averages, and concavity finishes the argument (End Matter). The relative gap $(G_{F,\infty}-E_{F,\infty})/E_{F,\infty}$ is thus positive for every $0<t<1$; the lower bound on it given by Theorem~\ref{thm:analytic} vanishes in the classical ($t\to0$) and pure ($t\to1$) limits and reaches $12.7\%$ near $t=0.62$.

\begin{figure}[tb]
\resizebox{0.9\columnwidth}{!}{%
\begin{tikzpicture}
\begin{semilogxaxis}[
  name=ga, width=0.98\columnwidth, height=4.4cm, log basis x=10,
  xmin=1, xmax=30, ymin=0.36, ymax=1.06,
  ylabel={ebits},
  xtick={1,2,5,10,20}, xticklabels={1,2,5,10,20},
  legend style={font=\footnotesize, at={(0.96,0.93)}, anchor=north east, draw=none, fill=none},
  tick label style={font=\footnotesize}, label style={font=\small}, clip=false]
\addplot[thick, blue] coordinates {(1,1.02346) (1.02,1.01756) (1.079,1.00101) (1.08,1.00060) (1.14,0.98468) (1.163,0.97877) (1.2,0.96965) (1.254,0.95668) (1.26,0.95541) (1.32,0.94184) (1.353,0.93464) (1.38,0.92888) (1.44,0.91644) (1.459,0.91256) (1.5,0.90442) (1.56,0.88888) (1.574,0.88518) (1.62,0.87270) (1.68,0.85630) (1.697,0.85150) (1.74,0.83965) (1.8,0.82305) (1.831,0.81493) (1.86,0.80735) (1.92,0.79253) (1.974,0.77981) (1.98,0.77853) (2,0.77405) (2.04,0.76532) (2.1,0.75284) (2.129,0.74697) (2.16,0.74104) (2.22,0.72988) (2.28,0.71933) (2.296,0.71652) (2.34,0.70933) (2.4,0.69987) (2.46,0.69091) (2.477,0.68848) (2.52,0.68240) (2.58,0.67434) (2.64,0.66668) (2.671,0.66284) (2.7,0.65941) (2.76,0.65250) (2.82,0.64592) (2.88,0.63967) (2.881,0.63957) (2.94,0.63371) (3,0.62803) (3.107,0.61853) (3.351,0.59961) (3.614,0.58265) (3.898,0.56748) (4.204,0.55392) (4.534,0.54179) (4.89,0.53095) (5,0.52799) (5.274,0.52125) (5.688,0.51254) (6.135,0.50472) (6.616,0.49766) (7.136,0.49130) (7.696,0.48554) (8.3,0.48032) (8.952,0.47557) (9.655,0.47126) (10,0.46939) (10.41,0.46733) (11.23,0.46374) (12.11,0.46047) (13.06,0.45747) (14.09,0.45473) (15.19,0.45221) (16.39,0.44990) (17.67,0.44779) (19.06,0.44584) (20,0.44468) (20.56,0.44405) (22.17,0.44240) (23.91,0.44089) (25.79,0.43949) (27.82,0.43820) (30,0.43702)};
\addlegendentry{$G_{F,\alpha}$}
\addplot[thick, black] coordinates {(1,0.99980) (1.02,0.99157) (1.079,0.96823) (1.08,0.96764) (1.14,0.94482) (1.163,0.93628) (1.2,0.92301) (1.254,0.90404) (1.26,0.90216) (1.32,0.88223) (1.353,0.87165) (1.38,0.86320) (1.44,0.84490) (1.459,0.83919) (1.5,0.82725) (1.56,0.81040) (1.574,0.80664) (1.62,0.79436) (1.68,0.77909) (1.697,0.77473) (1.74,0.76419) (1.8,0.75000) (1.831,0.74303) (1.86,0.73650) (1.92,0.72358) (1.974,0.71237) (1.98,0.71124) (2,0.70727) (2.04,0.69952) (2.1,0.68836) (2.129,0.68310) (2.16,0.67777) (2.22,0.66774) (2.28,0.65823) (2.296,0.65570) (2.34,0.64921) (2.4,0.64065) (2.46,0.63253) (2.477,0.63033) (2.52,0.62482) (2.58,0.61749) (2.64,0.61053) (2.671,0.60704) (2.7,0.60392) (2.76,0.59762) (2.82,0.59163) (2.88,0.58592) (2.881,0.58583) (2.94,0.58048) (3,0.57530) (3.107,0.56662) (3.351,0.54932) (3.614,0.53379) (3.898,0.51989) (4.204,0.50747) (4.534,0.49636) (4.89,0.48643) (5,0.48372) (5.274,0.47754) (5.688,0.46956) (6.135,0.46239) (6.616,0.45593) (7.136,0.45010) (7.696,0.44482) (8.3,0.44004) (8.952,0.43569) (9.655,0.43174) (10,0.43003) (10.41,0.42814) (11.23,0.42485) (12.11,0.42185) (13.06,0.41911) (14.09,0.41659) (15.19,0.41429) (16.39,0.41217) (17.67,0.41023) (19.06,0.40845) (20,0.40739) (20.56,0.40681) (22.17,0.40530) (23.91,0.40391) (25.79,0.40263) (27.82,0.40146) (30,0.40037)};
\addlegendentry{$E_{F,\alpha}$}
\addplot[blue, dashed, domain=1:30] {0.42245}; \addplot[black, dashed, domain=1:30] {0.38702};
\node[font=\small, anchor=north east, xshift=-14pt, yshift=-2pt] at (rel axis cs:0,0) {(a)};
\end{semilogxaxis}
\begin{semilogxaxis}[
  at={(ga.below south west)}, anchor=north west, yshift=-0.25cm,
  width=0.98\columnwidth, height=3.2cm, log basis x=10,
  xmin=1, xmax=30, ymin=0, ymax=11,
  xlabel={R\'enyi order $\alpha$}, ylabel={gap (\%)},
  xtick={1,2,5,10,20}, xticklabels={1,2,5,10,20},
  tick label style={font=\footnotesize}, label style={font=\small}, clip=false]
\addplot[thick, red] coordinates {(1,2.36628) (1.02,2.62068) (1.079,3.38618) (1.08,3.40619) (1.14,4.21842) (1.163,4.53776) (1.2,5.05293) (1.254,5.82342) (1.26,5.90183) (1.32,6.75764) (1.353,7.22676) (1.38,7.60859) (1.44,8.46734) (1.459,8.74380) (1.5,9.32804) (1.56,9.68318) (1.574,9.73578) (1.62,9.86243) (1.68,9.90999) (1.697,9.90908) (1.74,9.87395) (1.8,9.74121) (1.831,9.67605) (1.86,9.61932) (1.92,9.52829) (1.974,9.46680) (1.98,9.46122) (2,9.44185) (2.04,9.40725) (2.1,9.36664) (2.129,9.34996) (2.16,9.33389) (2.22,9.30583) (2.28,9.28183) (2.296,9.27587) (2.34,9.26134) (2.4,9.24388) (2.46,9.22903) (2.477,9.22529) (2.52,9.21642) (2.58,9.20573) (2.64,9.19670) (2.671,9.19255) (2.7,9.18907) (2.76,9.18264) (2.82,9.17725) (2.88,9.17272) (2.881,9.17265) (2.94,9.16894) (3,9.16579) (3.107,9.16144) (3.351,9.15574) (3.614,9.15327) (3.898,9.15251) (4.204,9.15253) (4.534,9.15282) (4.89,9.15313) (5,9.15321) (5.274,9.15337) (5.688,9.15353) (6.135,9.15362) (6.616,9.15367) (7.136,9.15369) (7.696,9.15370) (8.3,9.15371) (8.952,9.15371) (9.655,9.15371) (10,9.15371) (10.41,9.15371) (11.23,9.15371) (12.11,9.15371) (13.06,9.15371) (14.09,9.15371) (15.19,9.15371) (16.39,9.15371) (17.67,9.15371) (19.06,9.15371) (20,9.15371) (20.56,9.15371) (22.17,9.15371) (23.91,9.15371) (25.79,9.15371) (27.82,9.15371) (30,9.15371)};
\addplot[red!70!black, dashed] coordinates {(1,1.23100) (1.01,1.27428) (1.02,1.31454) (1.03,1.35179) (1.04,1.38601) (1.05,1.41720) (1.06,1.44536) (1.07,1.47049) (1.08,1.49257) (1.09,1.51159) (1.1,1.52755) (1.11,1.54044) (1.12,1.55026) (1.13,1.55698) (1.14,1.56061) (1.15,1.56113) (1.16,1.55853) (1.17,1.55281) (1.18,1.54394) (1.19,1.53193) (1.2,1.51675) (1.21,1.49840) (1.22,1.47687) (1.23,1.45214) (1.24,1.42421) (1.25,1.39305) (1.26,1.35867) (1.27,1.32104) (1.28,1.28016) (1.29,1.23601) (1.3,1.29219) (1.31,1.46173) (1.32,1.62958) (1.33,1.79573) (1.34,1.96017) (1.35,2.12289) (1.36,2.28387) (1.37,2.44309) (1.38,2.60055) (1.39,2.75624) (1.4,2.91014) (1.41,3.06223) (1.42,3.21251) (1.43,3.36097) (1.44,3.50759) (1.45,3.65235) (1.46,3.79526) (1.47,3.93629) (1.48,4.07543) (1.49,4.21267) (1.5,4.34801) (1.51,4.48142) (1.52,4.61291) (1.53,4.74245) (1.54,4.87004) (1.55,4.99566) (1.56,5.11931) (1.57,5.24098) (1.58,5.36065) (1.59,5.47832) (1.6,5.59397) (1.61,5.70761) (1.62,5.81921) (1.63,5.92877) (1.64,6.03628) (1.65,6.14173) (1.66,6.27043) (1.67,6.40293) (1.68,6.53312) (1.69,6.66100) (1.7,6.78655) (1.71,6.90976) (1.72,7.03062) (1.73,7.14913) (1.74,7.26527) (1.75,7.37903) (1.76,7.49042) (1.77,7.59941) (1.78,7.70600) (1.79,7.81018) (1.8,7.91196) (1.81,8.01131) (1.82,8.10824) (1.83,8.20274) (1.84,8.29480) (1.85,8.38442) (1.86,8.47160) (1.87,8.55632) (1.88,8.63860) (1.89,8.71841) (1.9,8.79577) (1.91,8.87067) (1.92,8.94310) (1.93,9.01306) (1.94,9.08056) (1.95,9.14558) (1.96,9.20814) (1.97,9.26823) (1.98,9.32584) (1.99,9.38098) (2,9.43365) (2.093,7.90046) (2.191,6.71574) (2.293,5.81472) (2.399,5.14491) (2.511,4.66326) (2.628,4.33412) (2.75,4.12784) (2.878,4.01961) (3.012,3.98875) (3.153,4.01801) (3.3,4.09312) (3.453,4.20233) (3.614,4.33609) (3.782,4.48670) (3.958,4.64807) (4.143,4.81543) (4.336,4.98513) (4.538,5.15444) (4.749,5.32138) (4.97,5.48453) (5.201,5.64295) (5.444,5.79605) (5.697,5.94348) (5.962,6.08511) (6.24,6.22091) (6.531,6.35097) (6.835,6.47542) (7.153,6.59445) (7.486,6.70824) (7.835,6.81701) (8.199,6.92096) (8.581,7.02029) (8.981,7.11521) (9.399,7.20590) (9.837,7.29257) (10.29,7.37537) (10.77,7.45450) (11.28,7.53010) (11.8,7.60234) (12.35,7.67137) (12.93,7.73732) (13.53,7.80034) (14.16,7.86055) (14.82,7.91809) (15.51,7.97307) (16.23,8.02560) (16.98,8.07579) (17.77,8.12375) (18.6,8.16958) (19.47,8.21336) (20.38,8.25520) (21.32,8.29518) (22.32,8.33338) (23.36,8.36988) (24.44,8.40475) (25.58,8.43808) (26.77,8.46992) (28.02,8.50034) (29.33,8.52941) (30,8.54346)};
\node[font=\small, anchor=north east, xshift=-14pt, yshift=-2pt] at (rel axis cs:0,0) {(b)};
\end{semilogxaxis}
\end{tikzpicture}}
\caption{\label{fig:alpha}(a) R\'enyi-$\alpha$ entanglement of formation of the normalized even sector of two Cooper pairs at $d=-\ln3$ as a function of the order $\alpha$, numerically optimized over Gaussian (blue) and all (black) decompositions; dashed lines mark the bounds of Theorem~\ref{thm:analytic} on $G_{F,\infty}$ (blue) and $E_{F,\infty}$ (black). (b) Relative gap $(G_{F,\alpha}-E_{F,\alpha})/E_{F,\alpha}$, numerically evaluated (solid) and analytically proven lower bound (dashed).}
\end{figure}
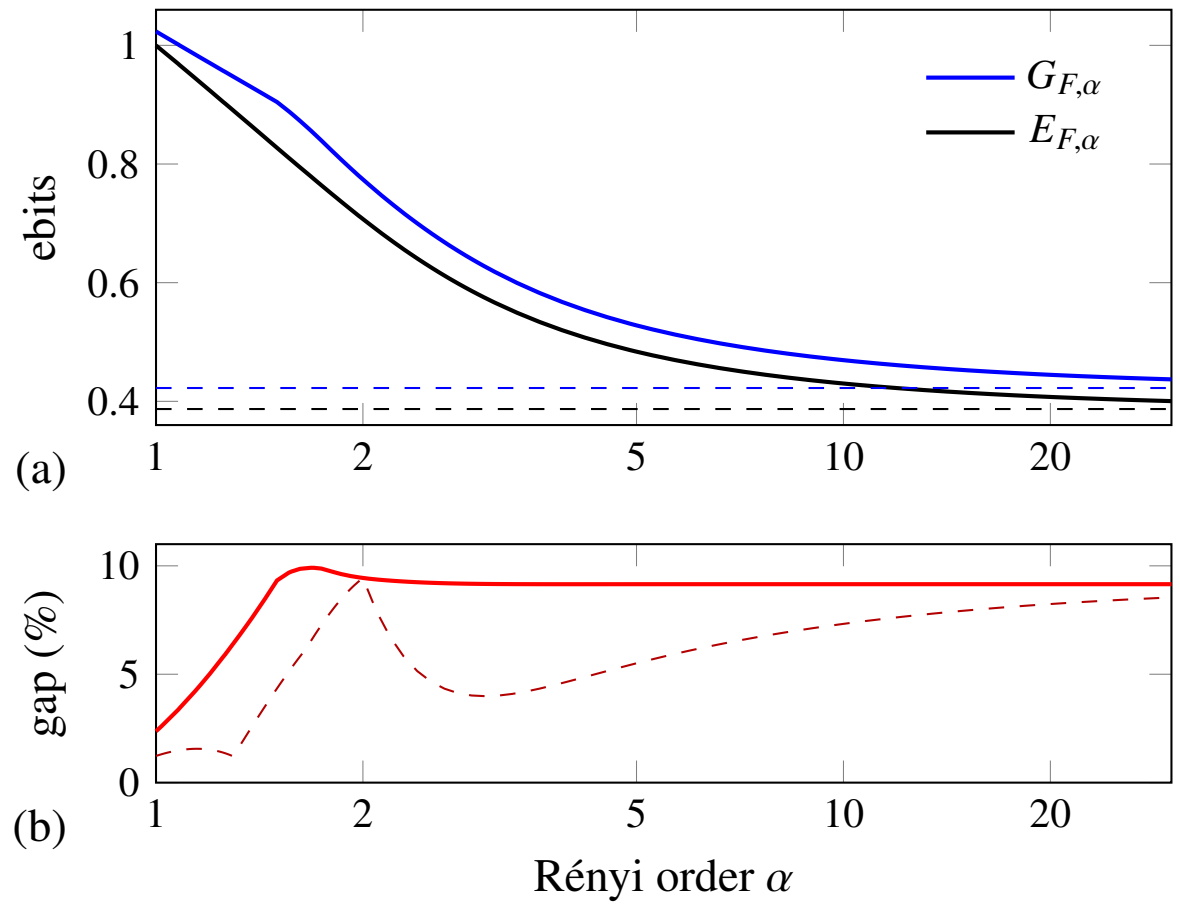

At finite order the whole Schmidt spectrum matters. We first derive two inequalities that hold for every pure Gaussian state, and then use them to settle, on one member of the isotropic family, Gaussian suboptimality for the entanglement of formation ($\alpha=1$) and, by interpolation, for all $\alpha>1$ (Fig.~\ref{fig:alpha}).

\begin{theorem}[All orders]\label{thm:renyi}
Every pure Gaussian state $\psi$ of two modes per party in the even sector satisfies
\begin{align}
 S(\psi_A)&\ge\tfrac{120}{17}\sqrt{\mathcal A\mathcal B}-\tfrac{30}{17}(\mathcal A+\mathcal B),
 \label{eq:lemma1}\\
 \|\psi_A\|_2&\le1+\tfrac{16}{9}(\mathcal A+\mathcal B)-\tfrac{40}{9}\sqrt{\mathcal A\mathcal B}.
 \label{eq:lemma2}
\end{align}
As a consequence, for the isotropic Cooper pair state $\rho$ at $t=\frac12$, i.e.\ $d=-\ln3$, $E_{F,\alpha}(\rho)<G_{F,\alpha}(\rho)$ for every $\alpha\ge1$.
\end{theorem}

The proof is in the Supplemental Material \cite{SM}; we sketch it here. A pure Gaussian state is, up to local Gaussian unitaries, a product of two pair states with concurrences $C_1,C_2$ \cite{BoteroReznik}, so its entropy is $\EW(C_1)+\EW(C_2)$. For given concurrences, $\mathcal A$ and $\mathcal B$ are bounded by explicit functions of $(C_1,C_2)$ (End Matter), and at these bounds the right-hand side of Eq.~\eqref{eq:lemma1} is bilinear in $(C_1,C_2)$; completing a square reduces Eq.~\eqref{eq:lemma1} to inequalities in one variable. At $\alpha=1$, averaging Eq.~\eqref{eq:lemma1} over a Gaussian decomposition and inserting Eq.~\eqref{eq:counting} gives $\GF(\hat\rho_+)\ge1.012$ ebits, whereas an explicit non-Gaussian decomposition, a mirror pair of states mixed with product states, achieves less than $1$ ebit. For $\alpha\ge2$, Eq.~\eqref{eq:lemma2} and Theorem~\ref{thm:analytic} suffice, because $\log\|x\|_\alpha$ is convex in $1/\alpha$. For $1<\alpha<2$ we interpolate: the R\'enyi entropy of a product of pair spectra is convex in $\alpha$, so its tangents at $\alpha=1$ and $\alpha=2$ are lower bounds affine in $\alpha$, and one further inequality for each tangent closes the interval. Every step is analytic (see Fig.~\ref{fig:alpha}).

\emph{Non-optimality in Schmidt number.}---The gap also shows up in a non-entropic quantity. The Schmidt number $\SN(\rho)$ is the least $r$ such that $\rho$ is a mixture of pure states of definite parity, as in Eq.~\eqref{eq:EF}, with Schmidt rank at most $r$. It cannot increase under LOCC, so preparing $\rho$ exactly by LOCC from a maximally entangled state requires one of Schmidt rank at least $\SN(\rho)$, i.e.\ at least $\log\SN(\rho)$ ebits. We denote by $\SN_G$ the Schmidt number with decompositions into pure Gaussian states, and by $\SN_{\rm q}$ the one with decompositions into arbitrary pure states of the Jordan--Wigner qubits, whose logarithm is the zero-error one-shot entanglement cost of $\rho$ under LOCC not restricted by superselection \cite{TerhalHorodecki,Nielsen,BuscemiDatta,YueChitambar}. Pure Gaussian states are, up to local Gaussian unitaries, products of independent pairs \cite{BoteroReznik}, so their Schmidt ranks are powers of $2$.

\begin{theorem}[Schmidt number]\label{thm:schmidt}
Let $\Pi_+$ project onto the even sector, $Y=\ketbra{\Phi_{\rm ee}}{\Phi_{\rm oo}}+\ketbra{\Phi_{\rm oo}}{\Phi_{\rm ee}}$ and $W=\ketbra{\Phi_{\rm ee}}{\Phi_{\rm ee}}+\ketbra{\Phi_{\rm oo}}{\Phi_{\rm oo}}$. Every mixture $\omega$ of pure states of definite parity and Schmidt rank at most $2$ satisfies $\Tr[(Y-\frac12\Pi_+)\omega]\le0$, and every mixture $\omega$ of pure Gaussian states of Schmidt rank at most $2$ satisfies $\Tr[(W-\frac12\Pi_+)\omega]\le0$. As a consequence, for the isotropic Cooper pair state $\rho$ at $t=\frac12$, i.e.\ $d=-\ln3$,
\begin{equation}
 \SN_{\rm q}(\rho)=2,\qquad\SN(\rho)=3,\qquad\SN_G(\rho)=4 .
\end{equation}
More generally, every physical state $\rho'$ with $D(\rho,\rho')<\frac3{128}$ has $\SN(\rho')\ge3$, and no state with $D(\rho,\rho')<\frac7{128}$ is a mixture of pure Gaussian states of Schmidt rank below $4$, where $D(\rho,\rho')=\frac12\|\rho-\rho'\|_1$ is the trace distance.
\end{theorem}

The witness inequalities, which give the lower bounds and the trace-distance radii, are proved in the End Matter \cite{TerhalHorodecki,SanperaBrussLewenstein,FriisCertification}; the matching upper bounds come from explicit decompositions, the flat one of Theorem~\ref{thm:analytic} for $\SN$ and one with $48$ elements for $\SN_{\rm q}$, whose remainder lies inside the separable ball around the maximally mixed state \cite{GurvitsBarnum,SM}. The two steps of this hierarchy have different origins. The step from $\SN_{\rm q}$ to $\SN$ is a superselection effect: the rank-two qubit decompositions superpose even and odd fermion number, just as the two-mode state $\tau_s=\frac14(\id-is\gamma_2\gamma_3)$ of Ref.~\cite{Salek}, with $s=1$, is separable as a two-qubit state but not under superselection \cite{BanulsCiracWolf}. The step from $\SN$ to $\SN_G$ is the Gaussian non-optimality gap exposed in this paper: both decompositions obey the same rules, and the Gaussian one needs one more unit of Schmidt rank as it cannot concentrate entanglement in one virtual pair.

Operationally, an LOCC protocol whose heralded branches, i.e.\ the pure states conditioned on all measurement outcomes, are all free fermion states needs two ebits to prepare $\rho$ exactly, whereas one ebit suffices on a qubit register with unrestricted LOCC \cite{TerhalHorodecki,Nielsen}.

\emph{How generic is the gap?}---More generally, let each Cooper pair be in the thermal state $\sigma=e^{-\mathcal H}/Z$ of the Hamiltonian
\begin{equation}
 \mathcal H=e(a^\dagger a+b^\dagger b)+d(a^\dagger b^\dagger+ba)+g(a^\dagger b+b^\dagger a),
 \label{eq:H}
\end{equation}
with mode energy $e$, pairing strength $|d|$ and hopping strength $|g|$; the isotropic family corresponds to $e=g=0$. 
We find that the gap persists without hopping ($g=0, e \ne0$), as certified by a state-independent Gaussian witness \cite{SM}; across the family without hopping, the relative gap at $\alpha=1$ reaches about $4\%$ at intermediate pairing. Hopping closes the gap, which vanishes numerically at $|g|=|d|$. The reason is the Gaussian lock: pairing couples $\ket{00}$ to $\ket{11}$ within each Cooper pair, and hopping couples $\ket{01}$ to $\ket{10}$. When the two are balanced, each Cooper pair feeds both virtual pairs equally, so the lock does not bind and a non-Gaussian decomposition has nothing to concentrate. The states $\tau_s$ of Ref.~\cite{Salek} are exactly the balanced thermal states of Eq.~\eqref{eq:H}, with $e=0$, $g=d$ and $s=\tanh|d|$, and we recover the Gaussian optimality proved there for two copies. With three copies, Gaussianity is no longer a single condition on two virtual pairs, and we find a gap numerically \cite{SM}.

The gap also reveals a second, non-Gaussian source of non-additivity. As $\GF=\EF$ on a single pair \cite{Salek},
\begin{equation}
 \begin{split}
 2\EF(\sigma)-\EF(\sigma^{\otimes2})=\;&\big[2\GF(\sigma)-\GF(\sigma^{\otimes2})\big]\\
 &+\big[\GF(\sigma^{\otimes2})-\EF(\sigma^{\otimes2})\big],
 \end{split} 
 \label{eq:split}
\end{equation}
where the first bracket is the Gaussian mechanism behind the strong non-additivity found by Salek \emph{et al.}\ and the second is the gap. For weak pairing $\GF$ is additive to numerical accuracy, and the entire non-additivity is non-Gaussian \cite{SM}.

\emph{A bound from second moments.}---For bosons, Gaussian optimality turns the Gaussian entanglement of formation into a lower bound on the entanglement of formation of every state with the same covariance matrix \cite{AdessoEoF,GA}. For fermions this route is closed, yet a bound from second moments can still be constructed. Let $\Delta=\sum_j(a_j^\dagger b_j^\dagger+b_ja_j)$ be the pairing operator of the two Cooper pairs, whose expectation value $\Tr(\rho\Delta)=2\sum_j\mathrm{Re}\langle a_j^\dagger b_j^\dagger\rangle$ is the total pairing amplitude \cite{KrausPairing}, which sets the BCS superconducting order parameter  \cite{BCS}.

\begin{theorem}[Second moments bound]\label{thm:pairing}
Every physical state $\rho$ of two modes per party, Gaussian or not, satisfies
\begin{equation}
 \EF(\rho)\ \ge\ 2\,\EW\big(|\Tr(\rho\Delta)|/2\big),
 \label{eq:pairbound}
\end{equation}
with equality for pure BCS states, i.e.\ products of two identical Cooper pair states $x\ket{00}+y\ket{11}$ with $x\bar y\ge0$.
\end{theorem}

The proof rests on an entropy inequality for classical bits: if $\lambda$ is a distribution on $N$ bits and $F_j$ its fidelity with the distribution obtained by flipping bit $j$, then $H(\lambda)\ge\sum_j\EW(F_j)$ (End Matter). Since $\Tr(\rho\,\mathcal U^\dagger\Delta\mathcal U)$ is linear in the covariance matrix for every local Gaussian unitary $\mathcal U$, and $\EF$ is invariant under such unitaries, maximizing the right-hand side over them gives a bound that depends on the covariance matrix alone. A measured pairing amplitude thus certifies entanglement in any state, and the certificate is exact on pure BCS states. It remains informative beyond them: for the non-Gaussian ground states of Eq.~\eqref{eq:H} with an added on-site interaction between the two Cooper pairs, the bound certifies more than half of their entanglement over a broad range of parameters \cite{SM}.

\emph{Discussion.}---We have shown that Gaussian decompositions of fermionic Gaussian states can be strictly suboptimal for entanglement quantification, proving the resulting gap analytically at every R\'enyi order $\alpha\ge1$ and in the Schmidt number. The gap stems from the interplay between superselection and Gaussianity: superselection splits two modes per party into two virtual qubit pairs, and Gaussianity locks their entanglement together. This contrasts sharply with bosonic Gaussian states, for which optimality of Gaussian decompositions has been recently established in full generality \cite{AdessoEoF,GA,GARenyi}.

The gap reflects an interplay between magic and entanglement. Matchgate circuits with classical randomness prepare exactly the mixtures of pure Gaussian states \cite{TerhalDiVincenzo,deMeloTerhal,BravyiKoenig}, and supplying any non-Gaussian pure state as a magic input makes these circuits universal \cite{Hebenstreit}. Our results show that magic can lower the entanglement needed to assemble a state that is itself free. For two modes per party, the magic required amounts to the violation of the Gaussian lock, quantified via the Williamson spectrum \cite{TurkeshiWilliamson}, or by other fermionic magic monotones \cite{DiasKoenig,ReardonSmith,CudbyStrelchuk}. Whereas magic can make entanglement manipulation computationally hard \cite{GuMagicSeparation}, we have shown that it can make the preparation of free fermion states cheaper.

Several questions are now open. The exact entanglement of formation of thermal Cooper pairs remains unknown, as does the dependence of the gap on the number of modes per party. The gap together with the non-additivity of Eq.~\eqref{eq:split} raises the question of whether the \emph{regularized} $\EF$ of fermionic Gaussian states may still be attained by Gaussian decompositions. In quantum chemistry, where orbital entanglement is increasingly quantified under parity superselection \cite{DingSchilling,DingMolecules,DingOrbitals}, our results imply that Gaussian decompositions of mean-field or BCS reduced states can overestimate their entanglement of formation when two orbitals per subsystem are paired across the bipartition. More broadly, superselection couples resources that would otherwise be independent, so that non-Gaussianity can substitute for entanglement even in assembling a free fermion state. Similar trade-offs may arise wherever symmetry constraints meet free dynamics, as with particle-number conservation in ultracold atoms or Gauss's law in lattice gauge theories \cite{BartlettRudolphSpekkens,BanulsLGT}.

\emph{Data availability.}---The code and certificates that support the findings of this article are openly available \cite{ZenodoCode}.

\begin{acknowledgments}
\emph{Acknowledgments.}---
This work was supported by the Engineering and Physical
Sciences Research Council (EPSRC) under Grant No.~UKRI6536. We acknowledge fruitful exchanges with Guillermo Garc\'ia-P\'erez, Tommaso Tufarelli, and Claudio Codice.
\end{acknowledgments}

\section*{End Matter}

\emph{Proof of Lemma~\ref{lem:quadric}.}---(i) In the Jordan--Wigner representation $c_j=Z_1\cdots Z_{j-1}\ketbra01_j$, with modes ordered as $a_1,a_2,b_1,b_2$, write an even state as $\ket\psi=\psi_0\ket0+\sum_{j<k}\beta_{jk}c_j^\dagger c_k^\dagger\ket0+\psi_4\,c_1^\dagger c_2^\dagger c_3^\dagger c_4^\dagger\ket0$. A Gaussian state with $\psi_0\ne0$ is proportional to $\exp(\frac12\sum_{jk}Q_{jk}c_j^\dagger c_k^\dagger)\ket0$ with $Q=\beta/\psi_0$ antisymmetric; on four modes the exponential stops at second order, and $\frac12\big(\frac12\sum_{jk}Q_{jk}c_j^\dagger c_k^\dagger\big)^2=\mathrm{Pf}(Q)\,c_1^\dagger c_2^\dagger c_3^\dagger c_4^\dagger$, so
\begin{equation}
 \psi_0\psi_4=\mathrm{Pf}\,\beta=\beta_{12}\beta_{34}-\beta_{13}\beta_{24}+\beta_{14}\beta_{23}.
\end{equation}
The manifold of even pure Gaussian states, $SO(8)/U(4)$, is connected and $\psi_0\ne0$ on a dense open subset, so the identity holds everywhere by continuity. In the Jordan--Wigner basis $\psi_0=\Psi_{00,00}$, $\psi_4=\Psi_{11,11}$, $\beta_{12}=\Psi_{11,00}$ and $\beta_{34}=\Psi_{00,11}$ are the entries of $\Pee$, and $\beta_{13}=\Psi_{10,10}$, $\beta_{24}=\Psi_{01,01}$, $\beta_{14}=\Psi_{10,01}$ and $\beta_{23}=\Psi_{01,10}$ those of $\Poo$, which turns the identity into $\det\Pee=-\det\Poo$. Conversely, on four modes the even pure spinors, i.e.\ the even pure Gaussian states, form exactly the nondegenerate quadric $\psi_0\psi_4=\mathrm{Pf}\,\beta$ in the projective even sector \cite{Chevalley}, so every even pure state with $\det\Pee+\det\Poo=0$ is Gaussian. More precisely, $|\psi_0\psi_4-\mathrm{Pf}\,\beta|=|\det\Pee+\det\Poo|$ is invariant under Gaussian unitaries, which act on $M$ by orthogonal conjugation and, by triality, on the even sector as $SO(8)$ on its vector representation; they therefore bring every even state to the form $z_0\ket0+z_4\,c_1^\dagger c_2^\dagger c_3^\dagger c_4^\dagger\ket0$, for which $M=(|z_0|^2-|z_4|^2)M_0$, with $M_0$ the covariance matrix of the vacuum. Hence $MM^\top=(1-4|z_0z_4|^2)\id=(1-4|\det\Pee+\det\Poo|^2)\id$. (ii) Alice's local Gaussian unitaries are generated by $\gamma_j\gamma_k$ with $j,k\le4$, which commute with her parity and therefore preserve her even subspace $\{\ket{00},\ket{11}\}$ and her odd subspace $\{\ket{01},\ket{10}\}$. The algebra $\mathfrak{so}(4)\simeq\mathfrak{su}(2)\oplus\mathfrak{su}(2)$ acts on these subspaces through its two commuting factors, so $\Pee\mapsto u_1\Pee$ and $\Poo\mapsto u_2\Poo$ with $u_k\in SU(2)$, and Bob's unitaries act likewise on the columns. A general parity-preserving local unitary acts as $U(2)\oplus U(2)$ and multiplies the two determinants by independent phases, so the states locally equivalent to Gaussian ones are those with $|\det\Pee|=|\det\Poo|$; all Gaussian bounds below use only this weaker property.\hfill$\blacksquare$

\emph{Counting statistics.}---The covariance matrix of $\Phi^+\otimes\Phi^+$ has vanishing local blocks and an orthogonal off-diagonal block, so every rotation $O_A$ of Alice's Majorana operators can be matched by a rotation of Bob's that leaves it invariant. The corresponding local Gaussian unitaries fix $\Phi^+\otimes\Phi^+$ and, since the covariance matrix of $\rho$ is $t$ times its own, also $\rho$; on the blocks they act as $\Pee\mapsto w_1\Pee\,\sigma_zw_1^\dagger\sigma_z$ and $\Poo\mapsto w_2\Poo w_2^\dagger$ with $w_k\in SU(2)$. Their commutant in the even sector is spanned by the $2\times2$ block on $\Phi_{\rm ee}=\sigma_z/\sqrt2$ and $\Phi_{\rm oo}=\id/\sqrt2$, in block coordinates, and by the identities on the two orthogonal triplets. The twirl of a pure state therefore depends only on $\mathcal P$, $\mathcal A$, $\mathcal B$ and the coherence $\langle\Phi_{\rm ee}|\Pee\rangle\overline{\langle\Phi_{\rm oo}|\Poo\rangle}$, of modulus $\sqrt{\mathcal A\mathcal B}$, and for $\rho$ one finds Eq.~\eqref{eq:counting}. A pure Gaussian state is, up to local Gaussian unitaries, a product of two pair states with concurrences $C_1,C_2$, by the fermionic modewise (BCS-like) decomposition \cite{BoteroReznik}, so its Schmidt coefficients are those of two independent bits. Its blocks have weights $\mathcal P=\frac12(1+\varepsilon \bar C_1\bar C_2)$ with $\bar C_j=\sqrt{1-C_j^2}$ and $\varepsilon=\pm1$, and von Neumann's trace inequality \cite{Mirsky} bounds its Bell-state probabilities by the squared sum of singular values of each block,
\begin{equation}
 \begin{split}
 \mathcal A&\le U=\tfrac14(1+\varepsilon \bar C_1\bar C_2+C_1C_2),\\
 \mathcal B&\le V=\tfrac14(1-\varepsilon \bar C_1\bar C_2+C_1C_2),
 \end{split}
 \label{eq:UV}
\end{equation}
with $\sqrt{UV}=\frac14(C_1+C_2)$. The Gaussian side of Theorem~\ref{thm:renyi} thus reduces to inequalities between $S_\alpha(\psi_A)$, a function of $(C_1,C_2)$ alone, and $\max_{\mathcal A\le U,\mathcal B\le V}[m\sqrt{\mathcal A\mathcal B}-\ell(\mathcal A+\mathcal B)]$, which is explicit \cite{SM}.

\emph{Proof of Theorem~\ref{thm:analytic}.}---By the counting statistics, $\Lambda$ and $\Lambda_G$ are suprema of $\mathbb E[\lambda_{\max}]$ over distributions of $(\mathcal P,\mathcal A,\mathcal B)$ obeying Eq.~\eqref{eq:counting}. For the lower bound on $\Lambda$, let $\mathcal P_\star=t^2/(1+t^4)$ and $\mathcal B_\star=(1+t^2)^2/[4(1+t^4)]$, and take an element with $\Pee=\sqrt{\mathcal P_\star}\,\Phi_{\rm ee}$, so that $\mathcal A=\mathcal P_\star$, and an odd block of weight $1-\mathcal P_\star$ with $\mathcal B=\mathcal B_\star$: for $t^2\ge2-\sqrt3$, where $\mathcal B_\star\ge(1-\mathcal P_\star)/2$, take $\Poo=\sqrt{\mathcal B_\star}\,\Phi_{\rm oo}+\sqrt{1-\mathcal P_\star-\mathcal B_\star}\,\Phi_{\rm oo}^\perp$ with $\ket{\Phi_{\rm oo}^\perp}=(\ket{01}\ket{01}-\ket{10}\ket{10})/\sqrt2$; otherwise take the rank-one block $\Poo=\sqrt{1-\mathcal P_\star}\,\ket v\bra{\bar w}$ with $|\langle \bar w|v\rangle|^2=2\mathcal B_\star/(1-\mathcal P_\star)$. This element and its mirror image under $\mathcal P\leftrightarrow1-\mathcal P$, $\mathcal A\leftrightarrow \mathcal B$, each with weight $\frac12$, have $\mathbb E[\mathcal P]=\frac12$, $(\mathcal P_\star+\mathcal B_\star)/2=a$ and $\sqrt{\mathcal P_\star \mathcal B_\star}=\kappa$, so their twirl reproduces $\hat\rho_+$, and all elements share one spectrum. The even block has eigenvalues $\frac{\mathcal P_\star}2,\frac{\mathcal P_\star}2$. In the first case the largest eigenvalue of the odd block, $\frac12(\sqrt{\mathcal B_\star}+\sqrt{1-\mathcal P_\star-\mathcal B_\star})^2$, simplifies to the stated bound, using $\sqrt{a\pm\kappa}=(1\pm t)^2/[2\sqrt{2(1+t^4)}]$ and $1-2a=3(1-t^2)^2/[4(1+t^4)]$; in the second the odd block has rank one and the largest eigenvalue is $1-\mathcal P_\star$.

For the Gaussian bound, let $\mathcal P_<=\min(\mathcal P,1-\mathcal P)$, $c$ the common determinant modulus and $y=\mathcal P_</2+c$. The overlap on the smaller block is at most $y$, which with $\delta=(\sqrt{\mathcal A}-\sqrt{\mathcal B})^2$ gives $\sqrt{\mathcal A\mathcal B}\le y+\sqrt{y\delta}$. Averaging, with $\mathbb E[\delta]\le2(a-\kappa)$ and Cauchy--Schwarz, yields $\kappa\le\mathbb E[y]+\sqrt{2(a-\kappa)\,\mathbb E[y]}$, i.e.\ $\mathbb E[y]\ge \mathcal P_\star$, since $2(a-\kappa)=(1-t)^4/[4(1+t^4)]$ and $2(a-\kappa)+4\kappa=(1+t)^4/[4(1+t^4)]$. At fixed $y$, $\lambda_{\max}=\frac12[1-\mathcal P_<+\sqrt{(1-\mathcal P_<)^2-4c^2}]$ increases with $c$, because $(1-\mathcal P_<)^2-4c^2=(1-2y)(1-2y+4c)$, so $\lambda_{\max}\le\phi(y)=\frac12(1-y+\sqrt{1-2y})$. As $\phi$ is concave and decreasing, $\mathbb E[\lambda_{\max}]\le\phi(\mathbb E[y])\le\phi(\mathcal P_\star)$, the stated bound.

For $t^2\in[2-\sqrt3,1)$ the difference of the two bounds is $2(1-t^2)[\sqrt3(1+t^2)-2\sqrt{1+t^4}]/[8(1+t^4)]>0$, since $3(1+t^2)^2-4(1+t^4)=-(t^4-6t^2+1)>0$ there; for $t^2<2-\sqrt3$, $\phi(\mathcal P_\star)<1-\mathcal P_\star$ because $\sqrt{1-2\mathcal P_\star}<1-\mathcal P_\star$. Since the flat decomposition gives $E_{F,\infty}(\hat\rho_+)\le-\log$ of the stated lower bound on $\Lambda$, and convexity of $-\log$ gives $G_{F,\infty}(\hat\rho_+)\ge-\log\Lambda_G$, the gap follows; the odd block cannot reverse it.\hfill$\blacksquare$

\emph{Proof of Theorem~\ref{thm:schmidt}.}---With $\Pi_+$, $Y$ and $W$ as in the theorem, in block variables $\langle Y\rangle_\psi=2\,\mathrm{Re}\big(\overline{\langle\Phi_{\rm ee}|\Pee\rangle}\langle\Phi_{\rm oo}|\Poo\rangle\big)\le2\sqrt{\mathcal A\mathcal B}$ for a parity-even $\psi$, and $\langle Y\rangle_\psi=\langle\Pi_+\rangle_\psi=0$ for a parity-odd one. If $\psi$ is even with Schmidt rank at most $2$ and both blocks nonzero, each block has rank one, so $\mathcal A\le \mathcal P/2$ and $\mathcal B\le(1-\mathcal P)/2$ and $2\sqrt{\mathcal A\mathcal B}\le\sqrt{\mathcal P(1-\mathcal P)}\le\frac12$; if one block vanishes, $\mathcal A\mathcal B=0$. Hence $Y-\frac12\Pi_+\le0$ on all parity-definite pure states of Schmidt rank at most $2$, and on their mixtures. Its spectrum, $\{\frac12,-\frac32,-\frac12,0\}$, lies in an interval of length $2$ and $\rho-\rho'$ is traceless, so $|\Tr[(Y-\frac12\Pi_+)(\rho-\rho')]|\le2D$ for states at trace distance $D$, and $\Tr[(Y-\frac12\Pi_+)\rho]=\frac{10}{32}-\frac{17}{64}=\frac3{64}$ gives $\mathrm{SN}(\rho')\ge3$ for $D<\frac3{128}$. For a pure Gaussian state, Schmidt rank at most $2$ forces $C_1C_2=0$ and, by Eq.~\eqref{eq:UV}, $\mathcal A+\mathcal B\le U+V=\frac12(1+C_1C_2)=\frac12$, so $W-\frac12\Pi_+\le0$ on such states; its spectrum, $\{\frac12,-\frac12,0\}$, lies in an interval of length $1$, and $\Tr[(W-\frac12\Pi_+)\rho]=\frac{41}{128}-\frac{17}{64}=\frac7{128}$. Since Gaussian Schmidt ranks are powers of $2$, any Gaussian decomposition of a state at trace distance less than $\frac7{128}$ contains elements of rank $4$. Conversely, at $t=\frac12$ the flat decomposition of Theorem~\ref{thm:analytic} has Schmidt rank $3$, and the odd block of $\rho$ is a mixture of products of a pair state with $\ket{01}$ or $\ket{10}$, of Schmidt rank at most $2$, so $\SN(\rho)\le3$.\hfill$\blacksquare$

\emph{Proof of Theorem~\ref{thm:pairing}.}---We first show the classical inequality: if $\lambda$ is a distribution of bits $z=(z_1,\ldots,z_N)$ and $F_j=\sum_z\sqrt{\lambda_z\lambda_{z\oplus e_j}}$, then $H(\lambda)\ge\sum_j\EW(F_j)$. Conditioning reduces entropy, so $H(\lambda)=\sum_jH(z_j|z_1\dots z_{j-1})\ge\sum_jH(z_j|z_{-j})$, with $z_{-j}$ all bits but the $j$th. For a single bit with distribution $(p,1-p)$ one has $h(p)=\EW(2\sqrt{p(1-p)})$ exactly, and $2\sqrt{p(1-p)}$ is the fidelity between $(p,1-p)$ and $(1-p,p)$. Writing $F(z_j|z_{-j})$ for this fidelity of the conditional distribution of $z_j$,
\begin{equation}
 \begin{split}
 H(z_j|z_{-j})&=\sum_{z_{-j}}p(z_{-j})\,\EW\big(F(z_j|z_{-j})\big)\\
 &\ge\EW\Big(\sum_{z_{-j}}p(z_{-j})F(z_j|z_{-j})\Big)=\EW(F_j),
 \end{split}
\end{equation}
by convexity of $\EW$. Now let $\psi$ be parity-even, with singular values $s_1,s_2$ of $\Pee$ and $r_1,r_2$ of $\Poo$, so that $\psi_A$ has spectrum $\lambda=(s_1^2,s_2^2,r_1^2,r_2^2)$. Assign the strings $00,11$ to $s_1^2,s_2^2$ and $01,10$ to $r_1^2,r_2^2$. Each string of even parity differs from each string of odd parity in exactly one bit, so $F_1+F_2=2\sum_{i,l}s_ir_l=2(s_1+s_2)(r_1+r_2)$. In block variables $\langle\Delta\rangle_\psi=2\,\mathrm{Re}\big[\overline{\Tr(\sigma_z\Pee)}\,\Tr\Poo\big]$, and the von Neumann trace inequality gives $|\Tr(\sigma_z\Pee)|\le s_1+s_2$ and $|\Tr\Poo|\le r_1+r_2$. Hence $|\langle\Delta\rangle_\psi|\le F_1+F_2$, and by the classical inequality, monotonicity and convexity of $\EW$,
\begin{equation}
 \begin{split}
 S(\psi_A)&=H(\lambda)\ge\EW(F_1)+\EW(F_2)\\
 &\ge2\,\EW\Big(\frac{F_1+F_2}{2}\Big)\ge2\,\EW\Big(\frac{|\langle\Delta\rangle_\psi|}{2}\Big).
 \end{split}
\end{equation}
For parity-odd $\psi$ the blocks are $\Psi_{\rm eo}$ (Alice even, Bob odd) and $\Psi_{\rm oe}$ (Alice odd, Bob even), and $\langle\Delta\rangle_\psi=2\,\mathrm{Re}\Tr\big(\Psi_{\rm eo}^\dagger\sigma_z\,\mathrm{adj}\Psi_{\rm oe}\big)$; since $\mathrm{adj}\Psi_{\rm oe}$ has the singular values of $\Psi_{\rm oe}$, $|\langle\Delta\rangle_\psi|\le2(s_1r_1+s_2r_2)\le F_1+F_2$ and the same chain applies.

For any decomposition $\rho=\sum_ip_i\psi_i$ into parity-definite pure states, $\sum_ip_iS(\psi_{i,A})\ge\sum_ip_i\,2\EW(|\langle\Delta\rangle_{\psi_i}|/2)\ge2\EW(|\Tr\rho\Delta|/2)$, again by monotonicity and convexity. Equality holds for $\psi=\chi\otimes\chi$ with $\chi=x\ket{00}+y\ket{11}$ and $x\bar y\ge0$, for which every inequality above is tight. Finally, for a local Gaussian unitary $\mathcal U$ the expectation $\Tr(\rho\,\mathcal U^\dagger\Delta\mathcal U)$ is a linear function of the covariance matrix of $\rho$, because $\mathcal U^\dagger\Delta\mathcal U$ is quadratic in the Majorana operators, and $\EF$ is invariant under $\mathcal U$; maximizing the bound over $\mathcal U$ therefore gives a lower bound that depends on the covariance matrix alone.\hfill$\blacksquare$

\bibliographystyle{apsrev4-2}
\bibliography{refs}

\clearpage
\onecolumngrid
\setcounter{page}{1}
\renewcommand{\thepage}{S\arabic{page}}
\setcounter{equation}{0}
\renewcommand{\theequation}{S\arabic{equation}}
\renewcommand{\theHequation}{S\arabic{equation}}
\setcounter{theorem}{0}
\renewcommand{\thetheorem}{S\arabic{theorem}}
\renewcommand{\theHtheorem}{S\arabic{theorem}}
\setcounter{table}{0}
\renewcommand{\thetable}{S\Roman{table}}
\renewcommand{\theHtable}{S\Roman{table}}
\setcounter{figure}{0}
\renewcommand{\thefigure}{S\arabic{figure}}
\renewcommand{\theHfigure}{S\arabic{figure}}
\setcounter{section}{0}
\setcounter{secnumdepth}{2}
\renewcommand{\thesection}{S\arabic{section}}

\begin{center}
{\large\bfseries Supplemental Material}\\[4pt]
{\large \papertitle}\\[4pt]
Gerardo Adesso
\end{center}

This Supplemental Material contains the proofs not given in full in the End Matter and the computations behind the examples of the Letter. Section~\ref{app:conventions} fixes conventions. Lemma~\ref{lem:quadric}, Theorem~\ref{thm:analytic}, Theorem~\ref{thm:schmidt} apart from its rank-two decomposition, and Theorem~\ref{thm:pairing} are proved in full in the End Matter. Section~\ref{app:renyi} proves Theorem~\ref{thm:renyi}, Sec.~\ref{app:schmidt} proves the Schmidt number without superselection in Theorem~\ref{thm:schmidt}, Sec.~\ref{app:generic} certifies gaps for thermal Cooper pairs with nonzero mode energy by a state-independent witness, Sec.~\ref{app:numerics} collects hopping, a family scan, non-additivity and R\'enyi orders, and Sec.~\ref{app:interacting} tests the second moments bound of Theorem~\ref{thm:pairing} on interacting, non-Gaussian states. Equation numbers without prefix refer to the Letter and its End Matter.

\section{Conventions and fermionic Gaussian states}\label{app:conventions}
This section collects the standard facts used throughout; see Refs.~\cite{SuraceTagliacozzo,HacklBianchi,BravyiLagrangian} for reviews and Refs.~\cite{BanulsCiracWolf,FriisLeeBruschi,SzalayFermionic} for mode entanglement under superselection.

\emph{Jordan--Wigner representation.}---We order the modes as $c_1=a_1$, $c_2=a_2$, $c_3=b_1$, $c_4=b_2$ and represent them on four qubits by
\begin{equation}
 c_j=Z_1\cdots Z_{j-1}\,\sigma^+_j,\qquad \sigma^+=\ketbra{0}{1},
 \label{eq:JW}
\end{equation}
where $Z_k$ is the Pauli $z$ operator on qubit $k$. Occupied modes are written as $\ket1$, so that
\begin{equation}
 c_j^\dagger c_k^\dagger\ket0=\ket{1_j1_k}\quad(j<k),\qquad c_1^\dagger c_2^\dagger c_3^\dagger c_4^\dagger\ket0=\ket{1111}.
\end{equation}
Because Alice's modes come first, her operators involve no string on Bob's qubits and are local in the qubit picture. Bob's operators of odd parity carry the string $Z_1Z_2$, i.e.\ Alice's parity, which commutes with every physical, parity-preserving reduced operator. Consequently, the fermionic partial trace over Bob's modes coincides with the ordinary partial trace over his qubits \cite{BanulsCiracWolf,FriisLeeBruschi}, and the Schmidt coefficients of a parity-definite state are those of the corresponding four-qubit state.

\emph{Majorana operators and covariance matrix.}---With the Majorana operators $\gamma_{2j-1}=c_j+c_j^\dagger$ and $\gamma_{2j}=i(c_j^\dagger-c_j)$ of the Letter, the covariance matrix of a state $\rho$ is
\begin{equation}
 M_{kl}=\tfrac i2\Tr\big(\rho[\gamma_k,\gamma_l]\big),\qquad k,l=1,\dots,8 .
\end{equation}
It is real and antisymmetric, and $\|M\|\le1$, i.e.\ $MM^\top\le\id$. A real orthogonal transformation $O$ of the Majorana operators brings it to the normal form
\begin{equation}
 M=O\Big[\bigoplus_{j=1}^4\begin{pmatrix}0&\lambda_j\\-\lambda_j&0\end{pmatrix}\Big]O^\top,\qquad 0\le\lambda_j\le1 .
 \label{eq:normalform}
\end{equation}

\emph{Gaussian states.}---A state is Gaussian if it is the thermal or ground state of a quadratic Hamiltonian, or a limit of such states. All its correlation functions then follow from $M$ by Wick's theorem, $\langle\gamma_{k_1}\cdots\gamma_{k_{2n}}\rangle=\mathrm{Pf}\big(-iM\big|_{k_1\dots k_{2n}}\big)$ for distinct indices \cite{BravyiLagrangian,SuraceTagliacozzo}. In the normal-mode Majoranas $\tilde\gamma=O^\top\gamma$, a Gaussian state factorizes as
\begin{equation}
 \rho=\prod_{j=1}^4\tfrac12\big(\id+i\lambda_j\tilde\gamma_{2j-1}\tilde\gamma_{2j}\big),
\end{equation}
so it is a mixture of the Fock states of its normal modes. These are pure Gaussian states, hence every Gaussian state has a Gaussian decomposition and $\GF(\rho)<\infty$. A normal mode is pure if $\lambda_j=1$ and mixed if $\lambda_j<1$; the number of mixed normal modes is the quantity used in Ref.~\cite{Salek} to classify the open cases. A pure state is Gaussian if and only if all $\lambda_j=1$, i.e.\
\begin{equation}
 MM^\top=\id .
\end{equation}

\emph{Gaussianity defect.}---For a pure state, $\|M\|_F^2=\Tr MM^\top\le8$, with equality exactly on Gaussian states. We therefore use the Gaussianity defect $8-\|M\|_F^2\ge0$ as a penalty in numerical optimizations over pure Gaussian states. For even states of two modes per party it is fixed by the Gaussian lock: by the identity proved in the End Matter, $8-\|M\|_F^2=32\,|\det\Pee+\det\Poo|^2$.

\section{Proof of Theorem~\ref{thm:renyi}}\label{app:renyi}
Theorem~\ref{thm:renyi} extends the gap of Theorem~\ref{thm:analytic} to every R\'enyi order $\alpha\ge1$. The Letter sketches the proof; here we give it in full.

\emph{Strategy.} The proof has three ingredients: four inequalities obeyed by every pure Gaussian element of the even sector, listed in Table~\ref{tab:lemmas}; their transfer to all R\'enyi orders, in two branches, $1\le\alpha\le2$ and $\alpha\ge2$; and two explicit non-Gaussian decompositions. Each inequality is linear in $\sqrt{\mathcal A\mathcal B}$ and $\mathcal A+\mathcal B$, so averaging it over a Gaussian decomposition and inserting $\mathbb E[\mathcal A+\mathcal B]=2a$, $\mathbb E\sqrt{\mathcal A\mathcal B}\ge\kappa$ gives the bound in the last column; for the tangent rows it applies at $\alpha=\frac{11}8$ and $\alpha=\frac{13}{10}$, because a tangent of the convex function $\alpha\mapsto S_\alpha$ lies below it, and for the $\alpha=2$ row it follows from convexity of $-\log$, as explained below. Throughout, $\log$ is to base $2$, $h_\alpha$ is the R\'enyi entropy of a pair, and a dot denotes $\partial_\alpha$.

\begin{table}[h]
\caption{\label{tab:lemmas}Inequalities for pure Gaussian elements, with $u=\sqrt{\mathcal A}$, $v=\sqrt{\mathcal B}$ and $\dot S_\alpha=\partial_\alpha S_\alpha(\psi_A)$. The third column is the condition for the extremum over $u\le\sqrt U$, $v\le\sqrt V$ to lie inside an edge; the last column is the bound on $G_{F,\alpha}(\hat\rho_+)$ at $d=-\ln3$ obtained by averaging, at the order $\alpha$ where each inequality is used.}
\begin{ruledtabular}
\begin{tabular}{llll}
role & inequality & edge & consequence\\ \hline
$\alpha=1$ & $S(\psi_A)\ge\frac{120}{17}uv-\frac{30}{17}(u^2+v^2)$, Eq.~\eqref{eq:lemma1} & $V\ge4U$ & $\GF\ge\frac{585}{578}$\\
tangent at $1$ & $S(\psi_A)+\frac38\dot S_1\ge-\frac3{50}+10uv-\frac{10}3(u^2+v^2)$ & $V\ge\frac94U$ & $G_{F,11/8}\ge\frac{1111}{1275}$\\
tangent at $2$ & $S_2(\psi_A)-\frac7{10}\dot S_2\ge-\frac3{200}+\frac{581}{50}uv-\frac{83}{20}(u^2+v^2)$ & $V\ge\frac{49}{25}U$ & $G_{F,13/10}\ge\frac{6123}{6800}$\\
$\alpha=2$ & $\|\psi_A\|_2\le1+\frac{16}9(u^2+v^2)-\frac{40}9uv$, Eq.~\eqref{eq:lemma2} & $V\ge\frac{25}{16}U$ & $G_{F,2}\ge2\log\frac{17}{13}$
\end{tabular}
\end{ruledtabular}
\end{table}

\emph{Gaussian elements.} A pure Gaussian $\psi$ has a Schmidt spectrum that is a product of two pair spectra with concurrences $C_1,C_2$, so every quantity in Table~\ref{tab:lemmas} is a function of one pair at a time, e.g.\ $S_\alpha(\psi_A)=h_\alpha(C_1)+h_\alpha(C_2)$ and $\|\psi_A\|_2^2=(1-\frac12C_1^2)(1-\frac12C_2^2)$. In the notation of Eq.~\eqref{eq:UV}, we write $\bar C_\times=\bar C_1\bar C_2$, $C_\times=C_1C_2$ and $C_+=C_1+C_2$, so that $\mathcal A\le U=\frac14(1+\varepsilon \bar C_\times+C_\times)$, $\mathcal B\le V=\frac14(1-\varepsilon \bar C_\times+C_\times)$, $\sqrt{UV}=\frac14C_+$ and $U+V=\frac12(1+C_\times)$; the blocks have determinant modulus $\frac14C_\times$.

\emph{Common reduction.} Each right-hand side is $\ell_0+muv-\ell(u^2+v^2)$, or its negative shifted by $1$, with $m>2\ell$. This quadratic form scales quadratically along rays from the origin, so its extremum over the box is at the corner $(\sqrt U,\sqrt V)$, on the axes, where the claims are immediate, or inside an outer edge. The edge $u=\sqrt U$ contains it only if $V\ge rU$ with $r=(m/2\ell)^2$; this forces $\varepsilon=-1$, and the extremal value is $\ell_0+\ell(r-1)U$. Writing $C_j=\sin\theta_j$ and $\theta_\pm=\theta_1\pm\theta_2$, one has $\bar C_\times-C_\times=\cos\theta_+$, so $U=\frac12\sin^2\frac{\theta_+}2$, and the edge condition reads $\cos\theta_-+r\cos\theta_+\ge r-1$. At fixed $\theta_+$ the allowed $\theta_-$ thus form a symmetric interval, and an edge claim $\varphi(\theta_1)+\varphi(\theta_2)\ge\Theta(\theta_+)$ reduces to $\theta_1=\theta_2$ as soon as $\omega=\varphi'$ obeys $\omega(\theta_1)\ge\omega(\theta_2)$ for $\theta_1\ge\theta_2$ in the region. For the three entropic inequalities the corner claim, $f(C_1)+f(C_2)\ge\ell_0+\frac m4C_+-\frac\ell2(1+C_\times)$, follows from $C_\times\ge sC_+-\frac12(C_1^2+C_2^2)-\frac{s^2}2$ (equality when $C_+=s$) and the one-variable inequality
\begin{equation}
 f(C)\ge\tfrac{\ell_0}2+\big(\tfrac m4-\tfrac{\ell s}2\big)C+\tfrac\ell4C^2-\tfrac\ell4+\tfrac{\ell s^2}8 .
 \label{eq:corner1}
\end{equation}

\emph{Exact verification.} The entropic one-variable inequalities involve logarithms. In $\upsilon=\tan\frac\theta2$, with $C=\sin\theta$ and $L=-\ln\upsilon$, one has $\ln2\,h_1=\ln(1+\upsilon^2)+\frac{2\upsilon^2}{1+\upsilon^2}L$, $\ln2\,\dot h_1=-\frac{2\upsilon^2}{(1+\upsilon^2)^2}L^2$, $\ln2\,h_2=2\ln(1+\upsilon^2)-\ln(1+\upsilon^4)$ and $\ln2\,\dot h_2=\ln(1+\upsilon^4)-\ln(1+\upsilon^2)+\frac{2\upsilon^4}{1+\upsilon^4}L$. On subintervals $[\upsilon_a,\upsilon_b]$ with rational endpoints we bound $\ln(1+x)$ by truncated series and $L$ by $-\ln\upsilon_b+\frac{2(\upsilon_b-\upsilon)}{\upsilon_b+\upsilon}\le L\le-\ln\upsilon_a-\frac{2(\upsilon-\upsilon_a)}{\upsilon+\upsilon_a}$, with rational enclosures of $\ln\upsilon_{a,b}$ and $\ln2$. Every check then becomes the positivity of a polynomial with rational coefficients on an interval, which we certify exactly by Sturm sequences or Bernstein coefficients. The verification code for this and every other certificate in this Supplemental Material is openly available \cite{ZenodoCode}, with a single script that reruns all checks. All margins are positive; the smallest, about $10^{-3}$ ebits, occurs in the corner inequality at $\alpha=\frac{13}{10}$.

\emph{Eq.~\eqref{eq:lemma2}.} At the corner the claim reads $9\sqrt{(2-C_1^2)(2-C_2^2)}\le T:=34+16C_\times-20C_+$. Here $T\ge10$, and $T^2-81(2-C_1^2)(2-C_2^2)=832-1360C_++562C_+^2+(764-640C_+)C_\times+175C_\times^2$ is convex in $C_\times$ on $[\max(0,C_+-1),C_+^2/4]$; it is positive at $C_\times=0$ and $C_\times=C_+-1$ (negative discriminants), equals $(5C-4)^2[T+9(2-C^2)]$ at $C_1=C_2=C$, and equals $\frac{162}{175}(-25C_+^2+40C_+-2)>0$ at its vertex, which is feasible only for $C_+\in[\frac{207}{145},1.502]$. On the edge the value is $1-U=\frac14(3+\bar C_\times-C_\times)$, and $(3+\bar C_\times-C_\times)^2-4(2+2\bar C_\times^2-C_\times^2)=(1-\bar C_\times-C_\times)(1+7\bar C_\times-5C_\times)\ge0$ because $(1-\bar C_\times)^2-C_\times^2=(\bar C_1-\bar C_2)^2$ and $C_\times\le\frac{16}{25}<\frac{52}{71}$ there.

\emph{Eq.~\eqref{eq:lemma1}.} Here $f=h_1=\EW$, $\ell_0=0$ and $s=\frac97$, so Eq.~\eqref{eq:corner1} reads $\EW(C)\ge\frac{75}{119}C+\frac{15}{34}C^2-\frac{15}{196}$; for $\upsilon\le\frac1{20}$ it follows from $\EW(C)\ge C^2$, i.e.\ $h(p)\ge4p(1-p)$. On the edge, $\theta_+\le\frac\pi3$, $\omega=\sin\theta\log\cot\frac\theta2$ increases up to $\theta_c\simeq0.58$, where $\cos\theta\ln\cot\frac\theta2=1$, and decreases after it, and a check of $\omega(\theta)\ge\omega(\theta'-\theta)$ for $\upsilon\in[0.27,\frac12]$, $\tan\frac{\theta'}2=\frac{231}{400}$, gives the symmetric reduction. At $\theta_1=\theta_2$ the claim is $\EW(C)\ge\frac{45}{34}C^2$ for $C\le\frac12$, which follows from $\EW(\frac12)=0.3546$ because $\EW(C)/C^2$ is non-increasing: in natural units and with $x=\sqrt{1-C^2}$, $2\EW-C\EW'=2\ln2-1-\sum_{n\ge1}\frac{2x^{2n}}{(2n-1)2n(2n+1)}\ge0$, since $\sum_{n\ge1}\frac1{(2n-1)2n(2n+1)}=\ln2-\frac12$.

\emph{Tangent at $\alpha=1$.} Here $f=h_1+\frac38\dot h_1$, $(\ell_0,m,\ell)=(-\frac3{50},10,\frac{10}3)$ and $s=\frac{47}{35}$; on the axes we use $f\ge-\frac1{40}$. On the edge, $\cos\theta_+\ge\frac19$ and $\cos\theta_j\ge\frac5{13}$. The derivative $\omega$ of $f(\sin\theta)$ has the sign of $\frac{11}8(1+\upsilon^2)-\frac38L(1-\upsilon^2)$, which increases with $\upsilon$, so $\omega$ changes sign once, at $\theta_0<0.06$; it is nondecreasing on $[\theta_0,2\arctan\frac25]$ and obeys $\omega(\theta)\ge\omega(\theta'-\theta)$ for $\upsilon\in[\frac25,\frac23]$, $\tan\frac{\theta'}2=\frac{179}{200}$. Together with $f\ge-\frac1{40}$, which settles $\sin^2\frac{\theta_+}2\le\frac1{250}$, this gives the symmetric reduction, and at $\theta_1=\theta_2$ the claim is $2f(C)\ge-\frac3{50}+\frac{25}{12}C^2$ for $C\le\frac23$.

\emph{Tangent at $\alpha=2$.} Here $f=h_2-\frac7{10}\dot h_2\ge h_2\ge0$, since $\dot h_2\le0$, which settles the axes; $(\ell_0,m,\ell)=(-\frac3{200},\frac{581}{50},\frac{83}{20})$ and $s=\frac{567}{400}$. On the edge, $\cos\theta_+\ge-\frac1{49}$ and $\cos\theta_j\ge\frac{12}{37}$. Now $f(\sin\theta)$ is convex for $\upsilon\le0.421$, beyond $\frac{\theta'}2$ with $\tan\frac{\theta'}2=\frac{49}{48}$, and $\omega(\theta)\ge\omega(\theta'-\theta)$ for $\upsilon\in[0.421,0.6]$; for $\upsilon_1\in[\frac35,\frac57]$ we use the edge condition, which decreases in both angles and vanishes at $(\upsilon_1,\upsilon_2)=(\frac35,\frac15)$: it forces $\upsilon_2\le\frac15$, and $\upsilon_2\le\frac18$ once $\upsilon_1\ge\frac{13}{20}$, while $\ln2\,\omega\ge\frac12$ on $[\frac35,\frac{13}{20}]$ and $\ge\frac25$ on $[\frac{13}{20},\frac57]$, and $\ln2\,\omega\le\frac12$ on $(0,\frac15]$ and $\le\frac25$ on $(0,\frac18]$; for $\upsilon\le10^{-3}$ the bound $\ln2\,\omega\le\frac{27}{10}\upsilon$ suffices. At $\theta_1=\theta_2$ the claim is $2f(C)\ge-\frac3{200}+\frac{249}{125}C^2$ for $C\le\frac57$.

\emph{Transfer to all orders.} Let $\zeta(\alpha)=\ln\sum_ip_i^\alpha$. Then $S_\alpha$ is minus the secant slope of $\zeta$ between $1$ and $\alpha$, and is convex in $\alpha$ wherever $\zeta'''$, the third cumulant of $\ln p$ under the escort distribution $p_i^\alpha/\sum_jp_j^\alpha$, is non-positive. For a pair this cumulant is $w(1-w)(1-2w)\ln^3\frac pq<0$, with escort weight $w>\frac12$ on the larger coefficient $p$. Hence for every Gaussian element $S_\alpha$ lies above its tangents at $\alpha=1$ and $\alpha=2$, both affine in $\alpha$. The $\alpha=2$ inequality gives $S_2=-2\log\|\psi_A\|_2\ge-2\log\frac{13}{17}-\frac{34}{13\ln2}(\|\psi_A\|_2-\frac{13}{17})$, by convexity of $-\log$, which is again linear in $\sqrt{\mathcal A\mathcal B}$ and $\mathcal A+\mathcal B$ and averages to $2\log\frac{17}{13}$. Interpolating linearly between the entries of Table~\ref{tab:lemmas},
\begin{equation}
 G_{F,\alpha}\ge\begin{cases}(1-\vartheta)\frac{585}{578}+\vartheta\frac{1111}{1275},&\alpha=1+\frac38\vartheta,\\[2pt] (1-\vartheta)\frac{6123}{6800}+\vartheta\,2\log\frac{17}{13},&\alpha=\frac{13}{10}+\frac7{10}\vartheta,\end{cases}
\end{equation}
with $\vartheta\in[0,1]$. For $\alpha\ge2$, since $\log\|x\|_\alpha$ is convex in $1/\alpha$, H\"older's inequality gives $\sum_ip_i\|\psi_{i,A}\|_\alpha\le\Lambda_G^{1-2/\alpha}(\frac{13}{17})^{2/\alpha}$, with $\Lambda_G\le\frac{13+3\sqrt{17}}{34}<\frac{13}{17}$ from Theorem~\ref{thm:analytic}; since $S_\alpha(\psi_A)=\frac\alpha{1-\alpha}\log\|\psi_A\|_\alpha$ and the logarithm is concave, $G_{F,\alpha}\ge[(\alpha-2)L_\infty+2\log\frac{17}{13}]/(\alpha-1)$ with $L_\infty=-\log\frac{13+3\sqrt{17}}{34}$.

\emph{Decompositions.} The decomposition at $\alpha=1$ takes $\mathcal P=\frac7{27}$, an element with $\Pee=\sqrt{\mathcal P}\,\Phi_{\rm ee}$, so $\mathcal A=\mathcal P$, and $\Poo=\sqrt{\mathcal B}\,\Phi_{\rm oo}+\sqrt{1-\mathcal P-\mathcal B}\,\Phi_{\rm oo}^\perp$ with $\mathcal B=\frac{25}{16}\mathcal P=\frac{175}{432}$. Its Schmidt coefficients are $\lambda=(\frac7{54},\frac7{54},\lambda_>,\lambda_<)$, $\lambda_\gtrless=\frac{160\pm5\sqrt{1015}}{432}$, and its entropy is $1.10163$ ebits. This element and its mirror image, each with weight $\frac{54}{119}$, together with $\ket{00}_A\ket{11}_B$ and $\ket{01}_A\ket{10}_B$, each with weight $\frac{11}{238}$, have $\mathbb E[\mathcal P]=\frac12$, $\mathbb E[\mathcal A]=\mathbb E[\mathcal B]=a$ and $\mathbb E\sqrt{\mathcal A\mathcal B}=\kappa$, so twirling turns them into a decomposition of $\hat\rho_+$ with average entropy $0.99981$ ebits. Its R\'enyi entropy is convex in $\alpha$: with $\Upsilon=\ln(\lambda/\lambda_>)\le0$ and $\mu_j=\mathbb E|\Upsilon|^j$ under the escort distribution, the third cumulant is $-\mu_3+3\mu_1\mu_2-2\mu_1^3\le\mu_2(3\mu_1-\delta)$, with $\delta=\ln\frac{54\lambda_>}7$ the smallest nonzero $|\Upsilon|$, and $\mu_1$ decreases with $\alpha$ from $0.4613<\frac\delta3=0.5803$. It therefore lies below its chords, which run from $0.99981$ to $0.86509$ on $[1,\frac{11}8]$ and from $0.88896$ to $0.71348$ on $[\frac{13}{10},2]$, below the two Gaussian lines. For $\alpha\ge2$ we use the flat decomposition of Theorem~\ref{thm:analytic}, whose elements all have spectrum $(\frac2{17},\frac2{17},\frac{13}{17})$ and $\alpha$-norm $(2^{\alpha+1}+13^\alpha)^{1/\alpha}/17>\frac{13}{17}$. The odd sector cannot reverse the inequality, since $E_{F,\alpha}\le G_{F,\alpha}$ there.\hfill$\blacksquare$

\section{Schmidt number without superselection}\label{app:schmidt}
Theorem~\ref{thm:schmidt} states that, if superselection is ignored and $\rho$ is regarded as a four-qubit state, it is a mixture of pure states of Schmidt rank $2$, $\SN_{\rm q}(\rho)=2$, whereas under superselection rank $3$ is needed. The End Matter proves the bounds under superselection; here we prove $\SN_{\rm q}(\rho)=2$.

\emph{Lower bound.} As a qubit state, $\sigma$ is Bell diagonal with weights $(\frac9{16},\frac1{16},\frac3{16},\frac3{16})$, so $\rho=\sigma\otimes\sigma$ has fidelity $(\frac9{16})^2>\frac14$ with the maximally entangled state of the two pairs, and $\SN_{\rm q}(\rho)\ge2$ \cite{TerhalHorodecki}.

\emph{Upper bound.} We set aside a small fraction of white noise, $\rho=(1-\eta)\tilde\rho+\eta\,\id/16$ with $\eta=10^{-3}$ and $\tilde\rho\ge0$, and decompose $\tilde\rho$ into $48$ vectors of Schmidt rank $2$. To find them, we maximize the weight carried by the two largest Schmidt coefficients over decompositions of $\tilde\rho$, by an iteration that increases it monotonically, and then truncate each vector to its two leading Schmidt terms. The resulting rank-two vectors reproduce $\tilde\rho$ up to rounding error. Whatever is left over, the noise $\eta\,\id/16$ plus this rounding residual, is therefore indistinguishable from a multiple of the maximally mixed state and lies deep inside the ball of separable states around $\id/16$, of radius $1/\sqrt{240}$ in Hilbert--Schmidt distance \cite{GurvitsBarnum}. Hence $\rho$ is a mixture of states of Schmidt rank at most $2$.
The rank-two elements found in this way superpose even and odd fermion number, with even-parity weights between $0.14$ and $0.86$, so they are not physical under superselection. This is the origin of the step from $\SN_{\rm q}=2$ to $\SN=3$.\hfill$\blacksquare$

\section{Thermal Cooper pairs with mode energy: a state-independent witness}\label{app:generic}
In the Letter (``How generic is the gap?'') we state that the gap persists for thermal Cooper pairs with nonzero mode energy, $e\ne0$, where the closed-form analysis of the isotropic family no longer applies. Here we prove this: a witness bounds $\GF$ from below for every state, explicit decompositions bound $\EF$ from above, and the two bounds are separated (Theorem~\ref{thm:main}).

\emph{A Gaussian witness.}---Let $\ket{\Phi^\pm}=(\ket{00}\pm\ket{11})/\sqrt2$ denote the Bell states of the Cooper pair $(a_j,b_j)$, write $\ket{\Phi^\pm\Phi^\pm}=\ket{\Phi^\pm}\otimes\ket{\Phi^\pm}$ for the corresponding state of both pairs, and consider
\begin{equation}
 X_\lambda=\lambda_+\ketbra{\Phi^+\Phi^+}{\Phi^+\Phi^+}+\lambda_-\ketbra{\Phi^-\Phi^-}{\Phi^-\Phi^-}.
 \label{eq:witness}
\end{equation}
In block variables, $\bra\psi X_\lambda\ket\psi=p(|\xi_3|^2+|\eta_0|^2)+q\,\mathrm{Re}(\bar\xi_3\eta_0)$ with $\xi_3=\Tr(\Pee\sigma_z)/\sqrt2$, $\eta_0=\Tr\Poo/\sqrt2$, $p=(\lambda_++\lambda_-)/2$ and $q=\lambda_+-\lambda_-$. Von Neumann's trace inequality \cite{Mirsky} bounds $|\xi_3|$ and $|\eta_0|$ by the singular values of the blocks, while Eq.~\eqref{eq:Sblocks} depends only on $\mathcal P$ and the two determinants. Imposing Lemma~\ref{lem:quadric}(i) therefore reduces the statement ``$S(\psi_A)\ge\bra\psi X_\lambda\ket\psi$ for all Gaussian $\psi$'' to an inequality $\Gamma_\lambda(\mathcal P,c)\ge0$ in two real variables, $\mathcal P\in[0,1]$ and the common determinant $c\in[0,\min(\mathcal P,1-\mathcal P)/2]$ (Sec.~\ref{app:witness}).

\begin{proposition}[State-independent Gaussian bound]\label{prop:witness}
Let $\mathcal K$ be the convex set of $(\lambda_+,\lambda_-)$ for which $\Gamma_\lambda\ge0$. Then for every state $\rho$,
\begin{equation}
 \GF(\rho_+)\ \ge\ \max_{\lambda\in\mathcal K}\ \big[\lambda_+F_{++}(\rho)+\lambda_-F_{--}(\rho)\big],
\end{equation}
with $F_{\pm\pm}(\rho)=\bra{\Phi^\pm\Phi^\pm}\rho\ket{\Phi^\pm\Phi^\pm}$. A rigorous branch-and-bound over $(\mathcal P,c)$ certifies, e.g., $(1.2329,-3.4823)\in\mathcal K$ up to a slack $10^{-5}$, with entropies in nats.
\end{proposition}

Without Lemma~\ref{lem:quadric}(i), i.e.\ allowing $|\det\Pee|\neq|\det\Poo|$, the same inequality is violated by $3.9\times10^{-2}$: this violation is precisely the room that non-Gaussian states enjoy. Adding the four-point coherence $\ketbra{0}{1111}+\mathrm{h.c.}$ and the projectors onto the local pair states $\ket{a_1a_2}$, $\ket{b_1b_2}$ to $X_\lambda$ gives a four-parameter witness that reduces in the same way and certifies the Gaussian bound also for strong pairing (Sec.~\ref{app:witness}). Because the reduction uses only $|\det\Pee|=|\det\Poo|$, all these bounds hold for the larger class of states that are locally unitarily equivalent to Gaussian ones: no local rotation of Gaussian elements, Gaussian or not, can close the gap.

\emph{Thermal Cooper pairs.}---Consider the thermal state $\sigma=e^{-\mathcal H}/Z$ of a Cooper pair, with $\mathcal H$ given by Eq.~\eqref{eq:H} at $g=0$, and the two-copy state $\rho=\sigma\otimes\sigma$ on the pairs $(a_1,b_1)$, $(a_2,b_2)$. For $d\ne0$ it has four mixed normal modes, the first case not covered by Ref.~\cite{Salek}. The single-copy value is exact: all parity-definite two-mode states are Gaussian and $\EF(\sigma)=p_{\rm e}\,\EW(C_{\rm e})$ with $p_{\rm e}$ the even-parity weight and $C_{\rm e}$ the concurrence of the even block \cite{Salek}.

\begin{theorem}[Non-optimality]\label{thm:main}
For the parameters in Table~\ref{tab:main}, $\GF(\sigma\otimes\sigma)-\EF(\sigma\otimes\sigma)$ is bounded below by the values listed there. In particular, Gaussian decompositions are not optimal for the entanglement of formation of fermionic Gaussian states.
\end{theorem}

\begin{table}[tbp]
\caption{\label{tab:main}Certified bounds for $\rho=\sigma\otimes\sigma$ in the even sector. Lower bounds on $\GF(\rho_+)$ come from Proposition~\ref{prop:witness} or its four-parameter extension, upper bounds on $\EF(\rho_+)$ from explicit 24-element ensembles with rigorous error control. All values in ebits.}
\begin{ruledtabular}
\begin{tabular}{cccc}
$(e,d)$ & $\GF(\rho_+)\ge$ & $\EF(\rho_+)\le$ & gap $\ge$\\ \hline
$(0,-0.8)$ & $0.356422$ & $0.342228$ & $1.4\times10^{-2}$\\
$(0.5,-1)$ & $0.480941$ & $0.465160$ & $1.5\times10^{-2}$\\
$(1,-1.1)$ & $0.537031$ & $0.517498$ & $1.9\times10^{-2}$\\
$(0,-1.2)$ & $0.606079$ & $0.597962$ & $8.1\times10^{-3}$\\
$(0,-1.5)$ & $0.805164$ & $0.802098$ & $3.0\times10^{-3}$\\
$(0,-2)$ & $1.150701$ & $1.150154$ & $5.4\times10^{-4}$
\end{tabular}
\end{ruledtabular}
\end{table}

The optimal non-Gaussian elements make the mechanism visible. At $(e,d)=(0,-0.8)$ the even-sector decomposition consists of product states and of entangled states that all share the reduced spectrum $(0.810,0.0945,0.0945,0.0007)$: they keep $\Poo$ maximally entangled while using the uncorrelated components $\ket{a_1a_2}$, $\ket{b_1b_2}$ of $\rho$ to make $\Pee$ almost singular. A Gaussian state with the same weights would be forced to carry $|\det\Pee|=|\det\Poo|$, and hence about $0.01$ in the smallest eigenvalue instead of $0.0007$.

\subsection{Reduction and certification}\label{app:witness}

\emph{Reduction.} Expanding $\Pee=\frac1{\sqrt2}\sum_\mu\xi_\mu\sigma_\mu$ and $\Poo=\frac1{\sqrt2}\sum_\mu\eta_\mu\sigma_\mu$ in Pauli matrices, one has $\|\Pee\|^2=\sum|\xi_\mu|^2=\mathcal P$ and $\det\Pee=\frac12(\xi_0^2-\xi_1^2-\xi_2^2-\xi_3^2)$. With $\ket{\pi_j}=a_j^\dagger b_j^\dagger\ket0$ and, in this section only, $\ket{1111}=a_1^\dagger b_1^\dagger a_2^\dagger b_2^\dagger\ket0$ in pair ordering, which is $-c_1^\dagger c_2^\dagger c_3^\dagger c_4^\dagger\ket0$, one has $\ket{\Phi^+\Phi^+}=\frac12(\ket0+\ket{\pi_1}+\ket{\pi_2}+\ket{1111})$, and a direct computation gives $\bra\psi X_\lambda\ket\psi=p(|\xi_3|^2+|\eta_0|^2)+q\,\mathrm{Re}(\bar\xi_3\eta_0)$. By the von Neumann trace inequality, $|\xi_3|\le(s_1+s_2)/\sqrt2=\sqrt{(\mathcal P+2c_{\rm e})/2}$ and $|\eta_0|\le\sqrt{(1-\mathcal P+2c_{\rm o})/2}$, where $s_i$ are the singular values of $\Pee$ and $c_{\rm e}=|\det\Pee|$, $c_{\rm o}=|\det\Poo|$. For Gaussian states $c_{\rm e}=c_{\rm o}=c$, so it suffices that
\begin{equation}
 \Gamma_\lambda(\mathcal P,c)=\mathcal S(\mathcal P,c)+\mathcal S(1-\mathcal P,c)-R_\lambda(\mathcal P,c)\ge0,\qquad
 R_\lambda=\max_{0\le x\le\sqrt{(\mathcal P+2c)/2},\ 0\le y\le\sqrt{(1-\mathcal P+2c)/2}}\big[p(x^2+y^2)+qxy\big],
\end{equation}
with $\mathcal S(x,c)$ the entropy of the eigenvalues $\frac12(x\pm\sqrt{x^2-4c^2})$. Since $p<0<q$ in the relevant range, $R_\lambda$ is attained at a corner of the box or at the vertex of the quadratic on an edge, and has a closed form.

\emph{Certification.} We cover the domain $\{0\le \mathcal P\le1,\ 0\le c\le\min(\mathcal P,1-\mathcal P)/2\}$ by an adaptive quadtree. On a cell $[\mathcal P_0,\mathcal P_1]\times[c_0,c_1]$ the larger eigenvalue $\frac12(x+\sqrt{x^2-4c^2})$ is increasing in $x$ and decreasing in $c$, the smaller one has the opposite monotonicity, and $-x\log x$ is concave, so its minimum over an interval is attained at an endpoint; this gives a rigorous lower bound on $\mathcal S(\mathcal P,c)+\mathcal S(1-\mathcal P,c)$. The box maximum $R_\lambda$ is monotone in the box size, which gives a rigorous upper bound. A cell is accepted when the difference exceeds $-\epsilon$ after subtracting a margin that comfortably covers floating-point rounding; otherwise it is split. For $(\lambda_+,\lambda_-)=(1.2329,-3.4823)$ and $\epsilon=10^{-5}$ the procedure terminates; certified constants and cell counts are listed in Table~\ref{tab:constants}. The bound in Proposition~\ref{prop:witness} then holds with $\lambda$ replaced by the certified constants and $\Tr(\rho_+X_\lambda)$ reduced by $\epsilon\Tr\rho_+$.

\emph{Four-parameter extension.} The optimal invariant dual for strong pairing has two further terms, $X=X_\lambda+\lambda_{\rm c}(\ketbra{0}{1111}+\ketbra{1111}{0})+\lambda_{\rm p}(\ketbra{a_1a_2}{a_1a_2}+\ketbra{b_1b_2}{b_1b_2})$. Then $\bra\psi X\ket\psi=(p+\lambda_{\rm c})|\xi_3|^2-\lambda_{\rm c}|\xi_0|^2+\lambda_{\rm p}(|\xi_1|^2+|\xi_2|^2)+p|\eta_0|^2+q\,\mathrm{Re}(\bar\xi_3\eta_0)$. For fixed moduli the phases of $\xi_0,\xi_1,\xi_2$ and of $\eta_1,\eta_2,\eta_3$ are free, which shows that $|\xi_0|^2,|\xi_3|^2\le(\mathcal P+2c)/2$ and $|\eta_0|^2\le(1-\mathcal P+2c)/2$ are the only constraints imposed by $|\det\Pee|=|\det\Poo|=c$. The inner maximum is a piecewise quadratic in $|\xi_3|$ and is computed exactly, so the same two-variable certification applies. The constants are obtained from a linear program whose constraints are linear in $(\lambda_+,\lambda_-,\lambda_{\rm c},\lambda_{\rm p})$; its value matches the numerically optimal Gaussian roof across the family.

\begin{table}[h]
\caption{\label{tab:constants}Certified witness constants (slack $\epsilon$), numbers of branch-and-bound cells (n.r.: not recorded), and resulting rigorous lower bounds on $\GF(\rho_+)$, with $\rho_+$ the unnormalized even block. The witness constants and slacks refer to entropies in nats, as used in the certification; the bounds are converted to ebits and rounded down.}
\begin{ruledtabular}
\begin{tabular}{ccccc}
$(e,d)$ & $(\lambda_+,\lambda_-,\lambda_{\rm c},\lambda_{\rm p})$ & $\epsilon$ & cells & $\GF(\rho_+)\ge$\\ \hline
$(0,-0.8)$ & $(1.2331,-3.5074,0,0)$ & $10^{-5}$ & $796\,752$ & $0.3564219$\\
$(0.5,-1)$ & $(1.23,-3.51,0,0)$ & $10^{-5}$ & $133\,126$ & $0.4809418$\\
$(1,-1.1)$ & $(1.2329,-3.4823,0,0)$ & $10^{-5}$ & n.r. & $0.5370315$\\
$(0,-1.2)$ & $(1.2332,-3.5122,0,0)$ & $10^{-5}$ & $741\,770$ & $0.6060797$\\
$(0,-1.5)$ & $(1.0904,-3.9089,0.4789,-0.4798)$ & $10^{-5}$ & $897\,923$ & $0.8051647$\\
$(0,-2)$ & $(0.9072,-4.3073,0.9423,-0.9425)$ & $3\times10^{-5}$ & $7\,525\,646$ & $1.1507016$
\end{tabular}
\end{ruledtabular}
\end{table}

\subsection{Explicit decompositions}\label{app:ensembles}

Upper bounds on $\EF$ are obtained by exhibiting decompositions. Writing a parity block as $\rho_s=\Omega\Omega^\dagger$ with $\Omega$ supported exactly on the sector, every decomposition is $\tilde\psi_i=\Omega v_i$ with $\sum_iv_iv_i^\dagger=\id$, i.e.\ a point of a Stiefel manifold. We minimize $\sum_i\|\tilde\psi_i\|^2S(\tilde\psi_i/\|\tilde\psi_i\|)$ with Riemannian conjugate gradients, using 24 elements per sector and several random restarts, and for $\GF$ we add a penalty proportional to the Gaussianity defect, which drives the elements to Gaussian states within numerical precision. The stored ensembles are then post-processed rigorously: the target is rebuilt from closed-form $2\times2$ blocks; for each element the reduced spectrum is enclosed by Weyl's inequality using the residual and the loss of orthogonality of the computed eigenvectors, and the entropy is bounded above on these enclosures by concavity of $-x\log x$; the reconstruction residual $\delta\rho=\rho_s-\sum_i\tilde\psi_i\tilde\psi_i^\dagger$ is bounded including floating-point error, and it is absorbed by writing $\rho_s=(1-r)\rho'+r\rho''$ with $r=\|\delta\rho\|/\lambda_{\min}$, which costs at most $r\log4$. These corrections are included in the bounds of Table~\ref{tab:ens}, and are far smaller than the gaps.

\begin{table}[h]
\caption{\label{tab:ens}Rigorous upper bounds from explicit ensembles, and the exact single-copy value, in ebits.}
\begin{ruledtabular}
\begin{tabular}{ccccc}
$(e,d)$ & $\EF(\rho_+)\le$ & $\EF(\rho_-)\le$ & $\EF(\sigma^{\otimes2})\le$ & $2\EF(\sigma)$\\ \hline
$(0,-0.8)$ & $0.3422273968$ & $0.2676946218$ & $0.6099220186$ & $0.6257187639$\\
$(0.5,-1)$ & $0.4651593990$ & $0.2892462722$ & $0.7544056711$ & $0.7789049921$\\
$(1,-1.1)$ & $0.5174974746$ & $0.2320528295$ & $0.7495503040$ & $0.7713672842$\\
$(0,-1.2)$ & $0.5979617472$ & $0.3516654160$ & $0.9496271632$ & $0.9884103590$\\
$(0,-1.5)$ & $0.8020971451$ & $0.3622840379$ & $1.1643811829$ & $1.2145244916$\\
$(0,-2)$ & $1.1501536189$ & $0.3146724467$ & $1.4648260655$ & $1.4985255115$
\end{tabular}
\end{ruledtabular}
\end{table}

\subsection{Symmetry reduction of the dual}\label{app:dual}

By convex duality, $\GF(\rho_+)=\sup\{\Tr(\rho_+X):\bra\psi X\ket\psi\le S(\psi_A)\ \text{for all pure Gaussian }\psi\}$. The family $\sigma\otimes\sigma$ is invariant under $u\otimes\bar u$, where $u\in U(2)$ rotates Alice's modes and $\bar u$ Bob's, and it is real in the Jordan--Wigner basis. Both symmetries are Gaussian and preserve entropies, so the optimal $X$ may be taken in the real commutant, which has dimension 9 in the even sector: a $3\times3$ block on the invariant vectors $\ket0$, $(\ket{\pi_1}+\ket{\pi_2})/\sqrt2$, $\ket{1111}$ and three multiples of the identity on the remaining isotypic components. The resulting nine-variable linear program, with constraints generated by a global search over the Gaussian manifold parametrized through Lemma~\ref{lem:quadric}(i), reproduces the numerically optimal Gaussian value and returns, for weak and moderate pairing, the operator $X_\lambda$ of Eq.~\eqref{eq:witness} with $\lambda\simeq(1.234,-3.50)$, nearly independent of $(e,d)$; this is what motivated Proposition~\ref{prop:witness}.

\section{Hopping, family scan, non-additivity and R\'enyi orders}\label{app:numerics}
This section gives the numerical evidence behind the last part of ``How generic is the gap?'' and behind Eq.~\eqref{eq:split} in the Letter: how hopping closes the gap, the gap for three copies, a scan over mode energies and pairing strengths, the split of non-additivity, and the gap at R\'enyi orders above one. Apart from the explicit bounds quoted for non-additivity, these results are numerical rather than certified.

\begin{figure}[tbp]
\resizebox{\columnwidth}{!}{%
\begin{tikzpicture}
\begin{axis}[
  name=pa, width=0.98\columnwidth, height=4.6cm,
  xlabel={$|d|$}, ylabel={ebits},
  xmin=0, xmax=2.6, ymin=0, ymax=0.053,
  legend style={font=\footnotesize, at={(0.02,0.98)}, anchor=north west, draw=none, fill=none},
  tick label style={font=\footnotesize}, label style={font=\small}, clip=false,
  yticklabel style={/pgf/number format/fixed}, scaled y ticks=false]
\addplot[thick, black, mark=*, mark size=1.2pt] coordinates {
 (0.2,0.00023) (0.4,0.00224) (0.6,0.00739) (0.8,0.01580) (1.0,0.02437) (1.2,0.03904) (1.5,0.05015) (2.0,0.03370) (2.5,0.01652)};
\addlegendentry{$2\EF(\sigma)-\EF(\sigma^{\otimes2})$}
\addplot[thick, blue, mark=square*, mark size=1.2pt] coordinates {
 (0.2,0.00023) (0.4,0.00224) (0.6,0.00739) (0.8,0.01464) (1.0,0.01235) (1.2,0.00954) (1.5,0.00395) (2.0,0.00075) (2.5,0.00013)};
\addlegendentry{$\GF(\sigma^{\otimes2})-\EF(\sigma^{\otimes2})$}
\addplot[thick, orange, dashed, mark=triangle*, mark size=1.4pt] coordinates {
 (0.2,0.00000) (0.4,0.00000) (0.6,0.00000) (0.8,0.00115) (1.0,0.01203) (1.2,0.02950) (1.5,0.04618) (2.0,0.03294) (2.5,0.01639)};
\addlegendentry{$2\GF(\sigma)-\GF(\sigma^{\otimes2})$}
\node[font=\small, anchor=north east, xshift=-14pt, yshift=-2pt] at (rel axis cs:0,0) {(a)};
\end{axis}
\begin{axis}[
  at={(pa.below south west)}, anchor=north west, yshift=-0.9cm,
  width=0.98\columnwidth, height=3.6cm,
  xlabel={$|g|/|d|$}, ylabel={$\GF-\EF$},
  xmin=0, xmax=1.25, ymin=0, ymax=0.0137,
  tick label style={font=\footnotesize}, label style={font=\small}, clip=false,
  yticklabel style={/pgf/number format/fixed}, scaled y ticks=false]
\addplot[thick, blue, mark=*, mark size=1.2pt] coordinates {
 (0.0,0.01234) (0.2,0.01133) (0.4,0.01128) (0.6,0.00579) (0.8,0.00046) (1.0,0.00000) (1.2,0.00022)};
\draw[densely dotted] (axis cs:1,0) -- (axis cs:1,0.0137);
\node[font=\small, anchor=north east, xshift=-14pt, yshift=-2pt] at (rel axis cs:0,0) {(b)};
\end{axis}
\end{tikzpicture}}
\caption{\label{fig:main}(a) Non-additivity of $\EF$ for two Cooper pair states at $e=0$ (black) and its exact split, Eq.~\eqref{eq:split}, into the non-optimality gap (blue) and the non-additivity of the Gaussian roof (orange). (b) Even-sector gap at $d=-1$ as hopping $g$ is added; it vanishes at balance $|g|=|d|$.}
\end{figure}
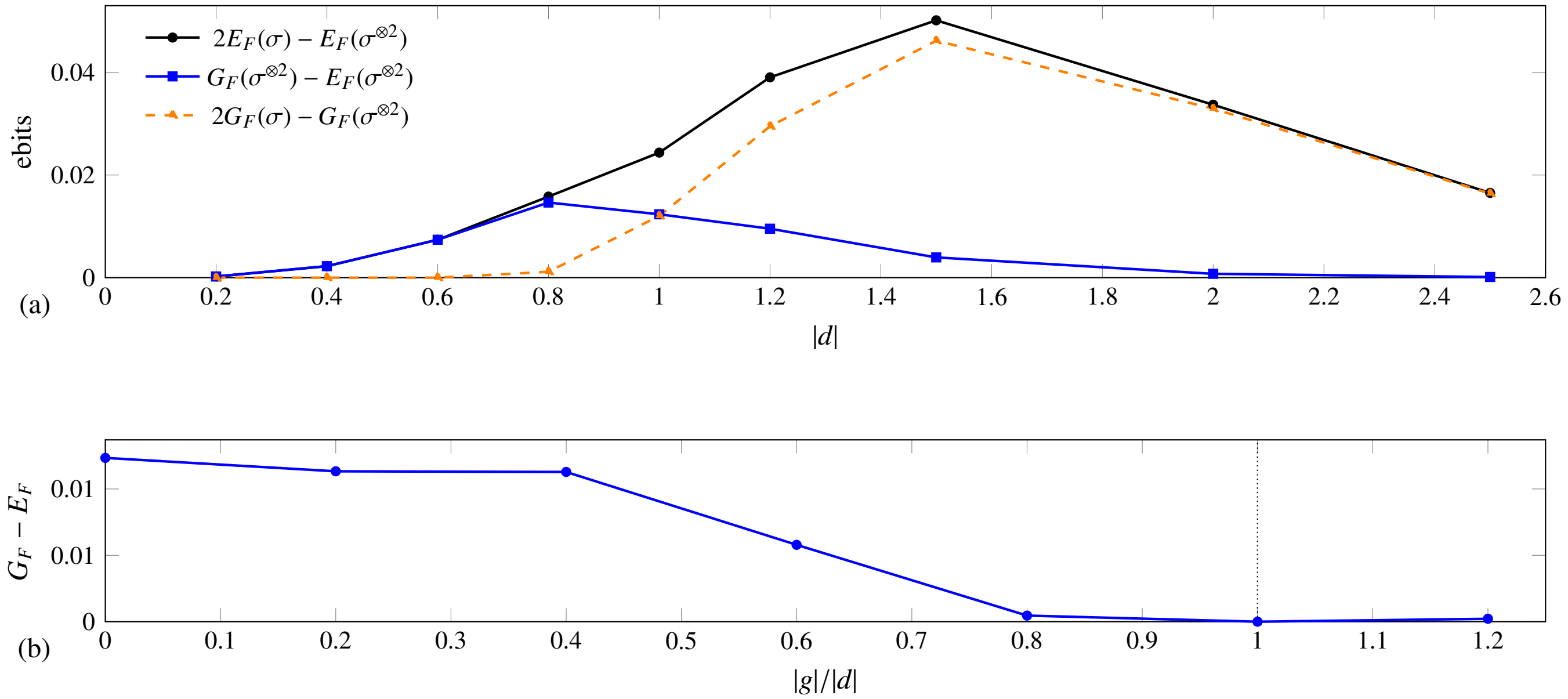

\emph{When does the gap appear?}---Turning on the hopping term $g(a^\dagger b+b^\dagger a)$ in Eq.~\eqref{eq:H} creates coherence in the odd sector of each pair, $\ket{01}\leftrightarrow\ket{10}$, alongside the pairing coherence $\ket{00}\leftrightarrow\ket{11}$. At $d=-1$ the even-sector gap decreases from $1.2\times10^{-2}$ at $g=0$ to $5.8\times10^{-3}$ at $g=-0.6$ and vanishes, within numerical accuracy, at the balanced point $|g|=|d|$ [Fig.~\ref{fig:main}(b)]. Balance is the situation in which the correlations of each pair feed both virtual qubit pairs equally, as for the two-copy states $\tau_s^{\otimes2}$ of Ref.~\cite{Salek} for which Gaussian decompositions were proven optimal; a particle--hole transformation of Bob's mode, which is a local Gaussian operation, exchanges pairing and hopping. Balance is, however, specific to two modes per party. For three copies of the same states, $\tau_s^{\otimes3}$ on three modes per party, the case left open in Ref.~\cite{Salek}, independent optimizations give a non-Gaussian advantage $\GF-\EF$ of $4.8\times10^{-3}$ at $s=0.4$, $7.2\times10^{-3}$ at $s=0.6$ and $3.5\times10^{-3}$ at $s=0.8$, equally split between the two parity sectors and confirmed by independent runs, while recovering exactly the proven values $\EF(\tau_1^{\otimes3})=2$ ebits and $\EF(\tau_s^{\otimes2})=\frac{1+s^2}{2}h\big(\frac{s^2}{1+s^2}\big)$. Numerically, then, Gaussian decompositions fail even for the very states on which their optimality was first established, as soon as a third copy is added.

\emph{Family scan.} We computed $\GF$ and $\EF$ for 40 states $\sigma_1\otimes\sigma_2$ with mode energies in $[-1.5,2]$, possibly different for the two modes of a pair, and $d\in[-2,-0.5]$, including unequal pairs. All show a positive even-sector gap, peaked at intermediate pairing $|d|\approx0.8$--$1.1$ and vanishing for weak and strong pairing; particle--hole symmetry $e\mapsto-e$ holds exactly, and the largest absolute gap is $1.98\times10^{-2}$ at $(e,d)=(\pm1,-1.1)$ and the largest relative gap $4.1\%$ at $(0,-0.8)$. Unequal pairs never exceed identical ones. Generic Gaussian states with four mixed modes show gaps of order $10^{-4}$ in both parity sectors, and no gap on a $1{:}3$ cut.

\emph{Non-additivity.}---Because $\GF(\sigma)=\EF(\sigma)$ for two modes, the non-additivity of $\EF$ splits exactly into two non-negative contributions, Eq.~\eqref{eq:split} of the Letter.
The first bracket is the mechanism behind the strong non-additivity found by Salek \emph{et al.}, $\EF(\tau_1^{\otimes m})=\lceil m/2\rceil$ ebits, where Gaussian ensembles entangled across copies are optimal \cite{Salek}. The second is new. Figure~\ref{fig:main}(a) shows that the two act in different regimes: for weak pairing the Gaussian roof is additive within numerical accuracy and all of the non-additivity is non-Gaussian, while for strong pairing the Gaussian mechanism takes over. With explicit ensembles we certify, for instance, $\EF(\sigma^{\otimes2})\le0.7495504$ against $2\EF(\sigma)=0.7713673$ at $(e,d)=(1,-1.1)$, and $\EF(\sigma^{\otimes2})\le1.1643812$ against $1.2145245$ at $(0,-1.5)$. These violations of additivity are modest next to those of Ref.~\cite{Salek}; their interest is that they arise from a different, genuinely non-Gaussian, mechanism.

\emph{R\'enyi orders.}---For bosons, Gaussian optimality and additivity hold for the R\'enyi entanglement of formation of every order $\alpha>1$ \cite{GARenyi}. For fermions neither is restored. Besides Theorems~\ref{thm:analytic} and~\ref{thm:renyi}, Table~\ref{tab:renyinum} shows that the even-sector gap for thermal Cooper pairs with nonzero mode energy persists at $\alpha=2,3,10$ and that $E_{F,\alpha}$ remains non-additive; in the odd sector we find no gap (data not shown).

\begin{table}[h] 
\caption{\label{tab:renyinum}Numerical R\'enyi gaps and non-additivity for $\sigma\otimes\sigma$ (unnormalized even block).}
\begin{tabular}{cccc}
\hline\hline
$\alpha$ & $(e,d)$ & $G_{F,\alpha}(\rho_+)-E_{F,\alpha}(\rho_+)$ & $2E_{F,\alpha}(\sigma)-E_{F,\alpha}(\sigma^{\otimes2})$ \\ \hline
2 & $(0,-0.8)$ & $1.0\times10^{-2}$ & $2.06\times10^{-2}$\\
2 & $(1,-1.1)$ & $1.79\times10^{-2}$ & $2.48\times10^{-2}$\\
3 & $(0,-0.8)$ & $7.5\times10^{-3}$ & $2.16\times10^{-2}$\\
3 & $(1,-1.1)$ & $1.4\times10^{-2}$ & $2.32\times10^{-2}$\\
10 & $(0,-0.8)$ & $5.6\times10^{-3}$ & $1.67\times10^{-2}$\\
10 & $(1,-1.1)$ & $1.0\times10^{-2}$ & $1.79\times10^{-2}$\\
\hline\hline
\end{tabular}
\end{table}

\section{Second moments bound on interacting states}\label{app:interacting}
To test Theorem~\ref{thm:pairing} beyond Gaussian states, we add to the pairing Hamiltonian of Eq.~\eqref{eq:H}, at $g=0$, an on-site interaction between the two modes of each party,
\begin{equation}
 H_J=d\,\Delta+e\sum_j(a_j^\dagger a_j+b_j^\dagger b_j)-J\big(a_1^\dagger a_1a_2^\dagger a_2+b_1^\dagger b_1b_2^\dagger b_2\big).
\end{equation}
The interaction couples the two Cooper pairs, so for $J\ne0$ the ground state is not Gaussian; its non-Gaussianity is measured by the flat Williamson spectrum $\nu=\sqrt{1-4|\det\Pee+\det\Poo|^2}$ (Lemma~\ref{lem:quadric}). For a pure state $\EF=S(\psi_A)$, so the bound can be compared with the exact value. At $d=-1$ we scanned $e\in[-2,2]$ and $J\in[-8,8]$ on a grid of $41\times81$ points; all $3321$ ground states are nondegenerate, have definite parity and carry $\EF>0.05$ ebits. Table~\ref{tab:interact} lists the fraction of $\EF$ certified by the bound $2\EW(|\langle\Delta\rangle|/2)$, evaluated in the given mode pairing without optimizing over local Gaussian unitaries. The fraction is never below $0.56$; it exceeds $0.93$ for all states with $\nu\ge0.99$ and decreases smoothly as the state departs from Gaussianity, with a median of $0.73$ for the most non-Gaussian states, $\nu<0.8$.

\begin{table}[h]
\caption{\label{tab:interact}Second moments bound on ground states of $H_J$ at $d=-1$: fraction $2\EW(|\langle\Delta\rangle|/2)/\EF$ of the exact entanglement certified, grouped by the Williamson value $\nu$.}
\begin{ruledtabular}
\begin{tabular}{cccc}
$\nu$ & states & minimum & median\\ \hline
$[0.54,0.8)$ & $358$ & $0.56$ & $0.73$\\
$[0.8,0.9)$ & $262$ & $0.78$ & $0.86$\\
$[0.9,0.95)$ & $216$ & $0.81$ & $0.92$\\
$[0.95,0.99)$ & $479$ & $0.84$ & $0.96$\\
$[0.99,1]$ & $2006$ & $0.93$ & $1.00$
\end{tabular}
\end{ruledtabular}
\end{table}

\end{document}